%% file: ms.tex
\documentclass{emulateapj}
\usepackage{subfigure}

\newcounter{countfig}
\newcounter{counttab}
\addtocounter{countfig}{1}
\addtocounter{counttab}{1}

\shorttitle{Highly-Obscured Star-Forming Regions in SINGS}
\shortauthors{Prescott et al.}

\begin{document}

\title{The Incidence of Highly-Obscured Star-Forming Regions in SINGS Galaxies}

\author{Moire K. M. Prescott\altaffilmark{1}, Robert C. Kennicutt, Jr.\altaffilmark{1,2,3}, 
George J. Bendo\altaffilmark{4,1}, Brent A. Buckalew\altaffilmark{5}, Daniela Calzetti\altaffilmark{3,6,7}, 
Charles W. Engelbracht\altaffilmark{1,3}, Karl D. Gordon\altaffilmark{1,3}, David J. Hollenbach\altaffilmark{8}, 
Janice C. Lee\altaffilmark{9,1,13}, John Moustakas\altaffilmark{10}, Daniel A. Dale\altaffilmark{3,11}, George Helou\altaffilmark{5}, 
Thomas H. Jarrett\altaffilmark{5}, Eric J. Murphy\altaffilmark{12}, John-David T. Smith\altaffilmark{1,3}, Sanae Akiyama\altaffilmark{1}, 
Megan L. Sosey\altaffilmark{6}}
\altaffiltext{1}{Steward Observatory, University of Arizona, 933 North Cherry Avenue, Tucson, AZ 85721; 
mprescott@as.arizona.edu, sakiyama@as.arizona.edu, cengelbracht@as.arizona.edu, kgordon@as.arizona.edu, jdsmith@as.arizona.edu}
\altaffiltext{2}{Institute of Astronomy, University of Cambridge, Madingley Road, Cambridge CB3 0HA, 
United Kingdom; robk@ast.cam.ac.uk}
\altaffiltext{3}{Visiting Astronomer, Kitt Peak National Observatory, National Optical Astronomy Observatories, 
which is operated by the Association of Universities for Research in Astronomy, Inc. (AURA) undercooperative agreement 
with the National Science Foundation.}
\altaffiltext{4}{Astrophysics Group, Imperial College, Blackett Laboratory, Prince Consort Road, London SW7 2AX United Kingdom; g.bendo@imperial.ac.uk}
\altaffiltext{5}{Caltech, IPAC, MS 220-6, 1200 E. California Boulevard, Pasadena, CA 91125;  brentb@ipac.caltech.edu, gxh@ipac.caltech.edu, jarrett@ipac.caltech.edu}
\altaffiltext{6}{Space Telescope Science Institute, 3700 San Martin Drive, Baltimore, MD 21218; calzetti@stsci.edu, sosey@stsci.edu}
\altaffiltext{7}{University of Massachussetts, Department of Astronomy, LGRT-B 619E, 710 North Pleasant Street, Amherst, MA 01003}
\altaffiltext{8}{NASA/Ames Research Center, MS 245-6, Moffett Field, CA 94035; hollenba@ism.arc.nasa.gov}
\altaffiltext{9}{National Optical Astronomy Observatory, 950 North Cherry Avenue, Tucson, AZ 85719; jlee@noao.edu}
\altaffiltext{10}{Department of Physics, New York University, 4 Washington Place, New York, NY 10003; john.moustakas@nyu.edu}
\altaffiltext{11}{Department of Physics and Astronomy, University of Wyoming, Laramie, WY 82071; ddale@uwyo.edu}
\altaffiltext{12}{Department of Astronomy, Yale University, P.O. Box 208101, New Haven, CT 06520; murphy@astro.yale.edu}
\altaffiltext{13}{Hubble Fellow.}

\begin{abstract}
Using the new capabilities of the Spitzer Space Telescope and extensive multiwavelength data from the 
Spitzer Infrared Nearby Galaxies Survey (SINGS), it is now possible to study the infrared properties of star formation 
in nearby galaxies down to scales equivalent to large H~\textsc{ii} regions.  We are therefore able to determine 
what fraction of large, {\it infrared-selected} star-forming regions in normal galaxies are highly obscured and address 
how much of the star formation we miss by relying solely on the optical portion of the spectrum.  
Employing a new empirical method for deriving attenuations of infrared-selected star-forming regions 
we investigate the statistics of obscured star formation on 500~pc scales in a sample of 38 
nearby galaxies.  We find that the median attenuation is 1.4 magnitudes in H$\alpha$ and that 
there is no evidence for a substantial sub-population of uniformly highly-obscured 
star-forming regions.  The regions in the highly-obscured tail of the attenuation distribution (A$_{H\alpha} \gtrsim$ 3) 
make up only $\sim$4\% of the sample of nearly 1800 regions, 
though very embedded infrared sources on the much smaller scales and lower luminosities of 
compact and ultracompact H~\textsc{ii} regions are almost certainly present 
in greater numbers.  The highly-obscured cases in our sample are generally the bright, central regions of 
galaxies with high overall attenuation but are not otherwise remarkable.  We also find that a majority of 
the galaxies show decreasing radial trends in H$\alpha$ attenuation.  The small fraction of highly-obscured 
regions seen in this sample of normal, star-forming galaxies suggests that on 500~pc scales the 
timescale for significant dispersal or break up of nearby, optically-thick dust clouds is short 
relative to the lifetime of a typical star-forming region.  

\end{abstract}

\keywords{galaxies: photometry -- galaxies: ISM --  stars: formation -- H~\textsc{ii} regions}

\section{INTRODUCTION}

Stars form enshrouded in their parent molecular clouds, but in time clusters of stars emerge from the obscuring dust 
and gas as H~\textsc{ii} regions and disperse part of this natal material.  
In the simplest case, the emission that we observe from such systems should undergo a transition from being strongly attenuated in 
the optical and ultraviolet (UV) to being less attenuated as the stars begin to break free of their birth clouds.  

In reality, the amount of obscuration observed in regions of star formation can vary widely depending on the environment.  
In nuclear starburst galaxies, V-band attenuations ($A_{V}$)\footnote{The attenuation at a given wavelength $\lambda$ 
is defined as $A_{\lambda} = k(\lambda) \times E(B-V)$ where $k(\lambda)$ is the assumed reddening curve and $E(B-V)$ is the color excess 
\citep[e.g.,][]{cal01}.  For comparison purposes, in this paper we assume $A_{H\alpha} = 0.82 \times A_{V}$.} 
as high as 10-100 magnitudes are commonplace.  The most well-studied of all, M82, has a central starburst with 
A$_{V}\sim$25-30 \citep{rie80, puxl91, matt01}, and 
the central starbursts of NGC~253 and NGC~4945 show $V$-band attenuations of 30-100 magnitudes \citep{genz98}.  
In a number of well-studied dwarf and irregular galaxies, a few compact 
regions dominate the infrared flux of the galaxy but remain invisible in the optical: Henize 2-10 \citep{kob99, vac02, cab05}, 
NGC~5253 \citep{gorj01}, SBS 0335-052 \citep{pla02, dal01, hou04}, and II~Zw~40 \citep{bec02}.  
\citet{soi01} found that more than half of the infrared luminosity in infrared-luminous starburst galaxies is generated 
in regions 100~pc--1~kpc in size, and
in the Antennae merger system, \citet{mir98} found that the most intense star-forming region is located outside both 
galaxy nuclei in an $\sim$50~pc area that is entirely unremarkable at optical wavelengths.

In contrast to these highly obscured cases, most normal H~\textsc{ii} regions show attenuations of only a few magnitudes.  
In the Large Magellanic Cloud, an $A_{V}\approx 1$ magnitude is typical \citep{cap86, bel02}.  
Balmer decrement measurements for optically-selected H~\textsc{ii} regions in samples of spiral galaxies 
yield attenuations of A$_{V}\sim$0.3-3.5 magnitudes \citep{zar94}.  The mean $V$-band attenuation derived from radio continuum measurements 
for H~\textsc{ii} regions in nearby galaxies is $\sim$1.7-2.0 magnitudes \citep{isr80, van88}.

Probing to smaller scales in the Milky Way, we see examples of a highly-embedded phase of star formation in the form of 
compact, ultracompact, or even ``superultracompact'' H~\textsc{ii} regions which range in size from 1~pc down to 0.01~pc \citep{kur00}. 
A study of the compact H~\textsc{ii} region S88B measured attenuations of A$_{V} \le 30$ along with 
attenuation gradients that suggested gaps in the molecular cloud had allowed optical emission to emerge alongside the thermal 
dust emission in the infrared \citep{goe03}.  
Results from carbon recombination line studies of ultracompact H~\textsc{ii} regions such as those in the W48 star-forming complex 
are consistent with being pressure-confined \citep{ros05a, ros05b}.  

The new capabilities of the Spitzer Space Telescope \citep{wer04} allow us to investigate the infrared properties of star formation within 
nearby galaxies down to scales as small as several hundred parsecs, the size of large H~\textsc{ii} regions.  
We are thus able to ask the following question: what fraction of {\it infrared-selected} star-forming regions in normal galaxies are highly 
obscured?  In other words, how much of the star formation in these galaxies have we been missing on these scales by only 
looking in the optical portion of the spectrum? 

Spitzer also gives us a new way to investigate radial attenuation properties of nearby galaxies.  Previous studies have 
found evidence for decreasing radial gradients in attenuation in samples of late-type galaxies using far-infrared to ultraviolet 
(FIR/UV) profiles and UV colors \citep{boi04, boi05, pop05, hil97} and in highly-inclined galaxies using broadband optical 
and near-infrared colors \citep{jan94}.  
In a detailed study of NGC~300, \citet{rou05} found a decreasing radial gradient in total 
infrared to far-ultraviolet (TIR/FUV).  They suggest that while variations in the metallicity, 
and presumably in the dust-to-gas ratio, may play a role in producing this gradient, 
other factors such as local variations in geometry or spectral energy distribution 
(SED) may be more important.  In contrast, \citet{zar94} found radial gradients in metallicity using optical spectra 
but saw only weak evidence for a gradient in attenuation.  With the advent of Spitzer, 
these questions can now be investigated on the scale of individual star-forming regions.

Taking advantage of the improved mid- and far-infrared resolution and sensitivity of Spitzer, the Spitzer Infrared Nearby 
Galaxies Survey (SINGS) is a multiwavelength imaging and spectroscopic survey of 75 galaxies within 30 Mpc \citep{ken03}.  
Designed to span the range of properties seen in nearby, normal galaxies, SINGS includes large spiral galaxies 
such as M51 and M81 as well as dwarf galaxies.  In addition to the 
Spitzer MIPS, IRAC, and IRS datasets, SINGS boasts a suite of ancillary data from radio continuum, HI, and CO to 
ground-based $BVRIJHK$, P$\alpha$, H$\alpha$, UV, and X-ray wavelengths.  
This combination of high resolution and multiwavelength coverage opens up the possibility of studying 
the attenuation properties of infrared-selected star-forming regions drawn from the local population of nearby galaxies.  

In this work, we investigated the attenuation properties of infrared-selected star-forming regions 
drawn from a sample of SINGS galaxies using multiband photometry and an empirical method for deriving attenuations of infrared-selected star-forming regions.
The paper is organized as follows.  We describe our galaxy sample in Section 2 and the observation and reduction steps in Section 3.  
In Sections 4 and 5 we describe our method of multiband photometry and our treatment of H$\alpha$ attenuation measurements.  
In Section 6 we present evidence of a paucity of large highly-obscured regions and of galaxy-wide radial trends in attenuation.  
We discuss the implications of these results for the lifetimes of star-forming regions in Section 7, and we conclude in Section 8.

\section{THE GALAXY SAMPLE}

The SINGS sample was selected to span the range of properties seen in nearby normal galaxies.  While not numerically representative of 
the local volume \citep{ken07b}, the survey was constructed to contain objects with the full range of morphological types, luminosities, and FIR/optical 
luminosity ratios \citep{ken03}.  The subset used in this work consists of 38 galaxies and represents nearly all of the 
galaxies in SINGS that have discrete infrared star-forming regions.  For reference, Table~1 gives the list of galaxies included in this 
work along with pertinent information.  The quoted $B$ magnitudes are total broadband 
magnitudes uncorrected for Galactic extinction and inclination and derive from the Third Reference 
Catalog of Bright Galaxies \citep[][hereafter RC3]{devauc91}.  
The galaxy dimensions were taken from \citet{ken03}; the morphological types were taken from RC3.  
The inclinations were calculated using the quoted galaxy dimensions and position angles taken from 
the 2MASS Large Galaxy Atlas \citep{jar03}, and the following standard equation \citep{hol58}:  
\begin{equation}
i = \cos^{-1}\left(\left(\frac{(b/a)^2-q_0^2}{1-q_0^2}\right)^{1/2}\right)
\label{eqn:incl}
\end{equation}   
where $a$ and $b$ are the optical major and minor axis diameters, and where $q_0$, the intrinsic optical axis ratio 
for an edge-on system, is set to 0.2 \citep{tul77}.  In the clearly edge-on cases -- NGC~3190 and NGC~4631 -- the inclination was set to 90$^{\circ}$, 
and the irregular galaxy NGC~6822 was set to an inclination of 0$^{\circ}$.  
While inclinations derived in this manner suffer from systematic errors due to warps and spiral arms, 
kinematically-derived inclination measurements exist for only a subset of the galaxies \citep[The HI Nearby Galaxies Survey, THINGS: ][]{wal05}.  
We have used the subset of our galaxy sample for which kinematically-derived inclinations exist in order to estimate the error 
in applying equation \ref{eqn:incl}.  The quoted error is the standard deviation of three measurements: the inclination derived from 2MASS ellipticities, 
an inclination derived by fitting an ellipse to the IRAC 3.6~$\mu$m data, and the kinematically-derived inclination 
from THINGS.  We quote the average error (7 degrees) for all other cases in which kinematically-derived inclinations are not available.  
The distances were taken from a number of sources with preference given to direct distance measurements from the literature; 
where direct measures were not available, we used flow-corrected estimates from \citet{mast07} (H$_{0} = 70$ km/s/Mpc).

\section{OBSERVATIONS AND REDUCTIONS}

The Spitzer images used in this work consist of the MIPS \citep{rie04} 24~$\mu$m and the IRAC \citep{faz04} 3.6, 4.5, 5.8, and 8.0~$\mu$m datasets obtained as part 
of the SINGS project.  The MIPS data reduction is described in \citet{gor05}, the IRAC data reduction in \citet{reg04}.  The 
ground-based H$\alpha$ and $R$-band images were obtained at NOAO using the KPNO 2.1-meter and CTIO 1.5-meter telescopes between 2001 and 2003.  
A description of the SINGS observing strategy can be found in \citet{ken03}.

Prior to beginning our analysis, the images were aligned, resampled to the same pixel scale, and convolved to match the full-width-half-maximum 
(FWHM) of the MIPS 24~$\mu$m point-spread-function (PSF) using SINGS\_TOOL\footnote{SINGS\_TOOL is an IDL program developed by Don Lindler 
(2006) for the SINGS Team for the automatic registration, alignment, resampling, and PSF-matching of multiwavelength images.}.  
Convolving to the 24~$\mu$m PSF mitigates some of the inconsistency that would result from 
comparing higher resolution ground-based optical data with lower resolution 24~$\mu$m imaging where some fraction of the light from a given 
point source has spilled into neighboring regions.  The 24~$\mu$m PSF, however, has fairly large diffraction lobes that are not replicated 
during our simple Gaussian convolution; for this reason, aperture corrections are applied at all wavelengths as described in Section 4.  

In order to correct the H$\alpha$ images for the stellar continuum contribution, we subtracted off a scaled R-band image.  
The resultant images were corrected for [N~\textsc{ii}]$\lambda\lambda$6548,6584\AA\ emission and the transmission curve of the H$\alpha$ filter; 
these methods are described in more detail in \citet{ken07b}.  
Where available, the applied [N~\textsc{ii}]$/$H$\alpha$ ratios were taken from SINGS radial driftscan spectra, where available \citep{mou06a}.  
Otherwise the corrections are approximate values derived from the $B$-band magnitude of the galaxy and the following empirical relation \citep{mou07}: 
\begin{equation}
\log([N~\textsc{ii}]\lambda6584]/H\alpha)= -5.19+0.468 Log(L_{B}/L_{B,\odot})
\label{eqn:niihalpha}
\end{equation}
\begin{equation}
\log([N~\textsc{ii}]\lambda6584]/H\alpha)= -0.187 (M_{B} - 22.254)
\label{eqn:niihalpha2}
\end{equation}
where $M_{B,\odot}=5.47$.  This relation is based on a sample of integrated spectra of galaxies \citep{mou06a, jan00} with the 
AGN component removed as described in \citet{mou06b}.  Table~2 summarizes the correction factors used.  

\section{REGION SELECTION AND MULTIBAND PHOTOMETRY}
Since most regions are unresolved at the 24~$\mu$m resolution ($\sim$6 arcseconds), 
we selected regions in the 24~$\mu$m image using DAOfind, a point-source selection task included in the DAOphot stellar 
photometry package \citep{ste87}.  For the initial selection we set a detection threshold of 10$\sigma$, which 
corresponds to a typical infrared luminosity limit of $\sim$3$\times10^{39}$ erg~s$^{-1}$); the infrared luminosity limits 
for each galaxy are given in Table~1.  For reference, 
the lowest corresponding H$\alpha$ luminosity limit is comparable to the luminosity of 
a few Orion-type H~\textsc{ii} regions ($\sim3\times10^{37} $erg s$^{-1}$).  
The median 24~$\mu$m luminosity of the region sample is $\sim$10$^{40}$, corresponding to a median H$\alpha$ luminosity 
of $\sim$2$\times$10$^{38}$ erg s$^{-1}$ or a few tens of Orion-type H~\textsc{ii} regions.  

\begin{figure}
\plotone{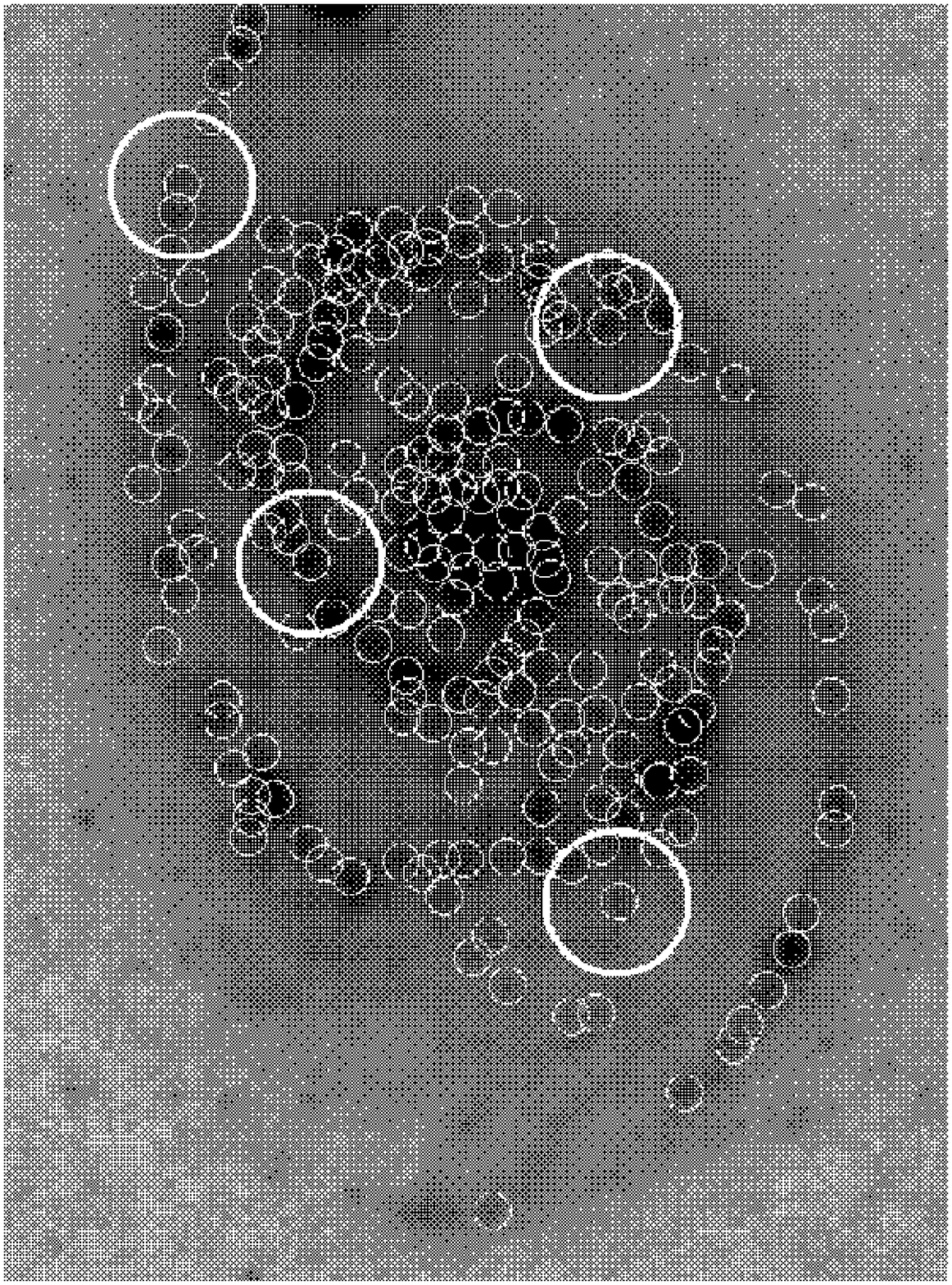}
\caption{24~$\mu$m image of NGC~5194 (M51) with the final region selection indicated as small black circles, each 500~pc in diameter, 
along with several examples of background annuli (white circles), each with an inner diameter of 500~pc and an outer diameter of 2000~pc.}  
\label{fig:m51}
\end{figure}

We measured region fluxes using 500~pc diameter circular apertures ($\sim$ 13\arcsec\ for a galaxy at 8~Mpc).  
While this aperture size is larger than a typical H~\textsc{ii} region ($\sim$100~pc), it was chosen as a compromise 
in order to accommodate the large 24~$\mu$m PSF and to allow a comparison of regions of the same physical 
size between galaxies at different distances.  As an example, in Figure~\ref{fig:m51} we show the 24~$\mu$m image of NGC~5194 with the final 
region selection of 500~pc circular regions marked.  In some cases, a 500~pc region was smaller than the FWHM of 
the 24~$\mu$m PSF; the aperture corrections discussed below were used to address this issue.

Measuring robust fluxes for this work was complicated by the issue of how 
to determine the background value that should be subtracted from the raw region flux measurement.  In addition to the sky 
background of the image, each 24~$\mu$m source sits on top of a diffuse component associated with the overall galaxy profile 
and internal structures such as spiral arms.  Simple aperture photometry methods dictate using an annulus around each aperture 
to get a measure of the local background; however at the 24~$\mu$m resolution many star-forming regions lie within what would 
make up the sky annulus of a neighboring region, thereby contaminating the background estimate.  On the other hand, ignoring 
the contribution of the diffuse component and subtracting only a global sky background is not entirely satisfactory either as 
it systematically overestimates the flux by 1--10\% for regions on spiral arms or near the center of the galaxy.    

For this study, therefore, we employed three separate methods of sky subtraction.  First, we used a global sky subtraction 
method by subtracting a simple sky background defined as the $\sigma$-clipped mean value from two off-galaxy portions of the image.  
Second, we employed a local background subtraction method where we measured the background within large annuli (spanning 
from the measurement aperture to an outer radius equal to 4 times the measurement aperture radius) 
centered on each region using a $\sigma$-clipped mode or mean estimation technique.  The large
annuli are needed to minimize contamination due to the inclusion of neighboring IR-selected regions.  To 
further reduce contamination, we removed pixels assigned to neighboring IR-selected regions prior to calculating the local background value.  
The success of this technique depends on the size of the 
background annuli used as well as properties of the galaxy such as the inclination, the spacing of the spiral arms, and the 
surface density of star-forming regions across the face of the galaxy.  Figure~\ref{fig:m51} shows a small subset of the background annuli used 
in the case of NGC~5194.  Finally, we applied a profile background subtraction method which involved dividing the galaxy into 
elliptical annuli according to the inclination and position angle and deriving a $\sigma$-clipped mean background value at each 
radius.  As in the local background subtraction method, we removed pixels belonging to IR-selected regions prior to calculating the 
background value in order to reduce the effects of contamination.  This method is less sensitive to neighbor contamination than 
the local background method, but it does assume azimuthal symmetry.  Fortunately, a majority of the targets in our 
current study are spiral galaxies that do not deviate far from this assumption.    

\begin{figure}
\plotone{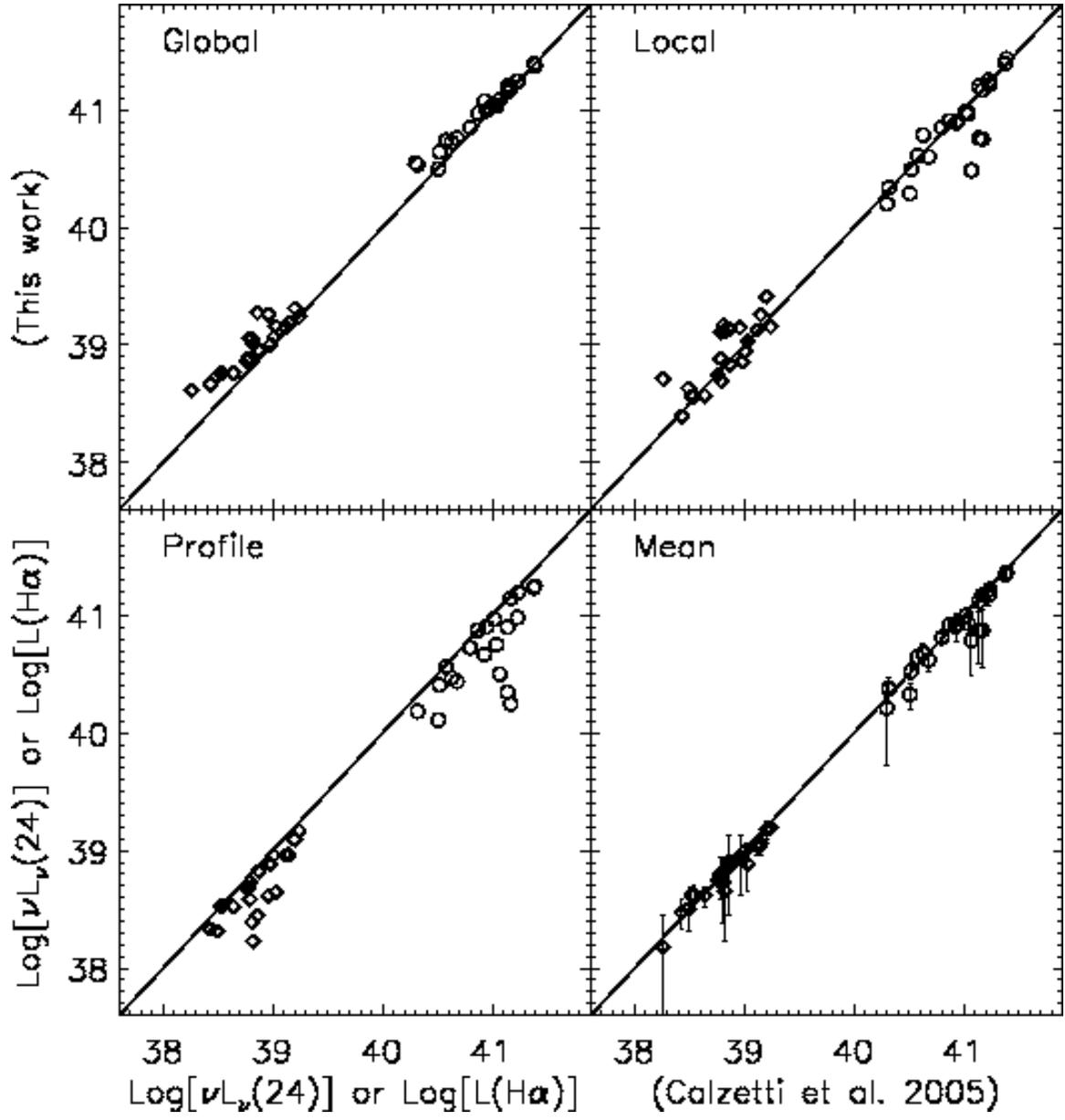}
\caption{Measurements of region luminosities - $\nu L_{\nu}(24)$ (circles) and $L(H\alpha)$ (diamonds) - for the 
four different techniques described in the text 
compared with those of \citet{cal05} for a sample of matched 13 arcsecond ($\sim$517~pc) regions in NGC~5194.  Final flux measurements were made using the 
mean of the global, local, and profile methods, as described in Section~4.}
\label{fig:calzetticompare}
\end{figure}

As an independent check on our background methods, we compared our flux measurements with those of \citet{cal05}, a detailed 
multiwavelength study of NGC~5194.  In that study background subtraction of region flux measurements was done using many rectangular 
background areas defined by hand with statistics computed on the portion of the background area appropriate to the region being 
measured.  We make the assumption that the level of fine-tuning demonstrated in \citet{cal05} for NGC~5194 provides the closest 
approximation of the true local background for each region that one could achieve in treating each galaxy by hand.  In the first 
three panels of Figure~\ref{fig:calzetticompare} we show a comparison between the 24~$\mu$m and H$\alpha$ luminosities measured with each of the three 
background methods versus those of \citet{cal05} for a matched sample of 13 arsecond ($\sim$517~pc) regions in NGC~5194.  
We find that the most luminous regions are measured consistently regardless of the background subtraction method because the 
background flux makes up a very small fraction of the raw region measurement.  For faint, low surface brightness regions, however, 
the background is a more considerable portion of the measurement and thus the final value is highly dependent on the method applied.  
The global sky subtraction systematically underestimates the background because it neglects the diffuse component 
of the galaxy.  The local background subtraction using large sky apertures is robust in many cases but tends to fail in crowded areas.  
The profile background subtraction tends to overestimate the background in interarm regions and underestimate 
the background in the spiral arms.  
Taken together, there is no standard method for dealing with the complexity of these nearby galaxies that is not in some 
way biased by the assumptions made or by the manner in which it is applied.  In this work, we have taken the average of these 
three methods as our best estimate of the flux and quote a standard error of the mean.  In doing so, we reject 
individual measurements that are negative (i.e., the background was estimated to be higher than the source) as unphysical and 
average the remaining two measurements; this occured in only a small fraction ($\sim$5\%) of the region sample.  
The success of the average versus the results of \citet{cal05} is shown in the last panel 
of Figure~\ref{fig:calzetticompare}.

We estimated aperture corrections using the available MIPS and IRAC PSFs\footnote{http://ssc.spitzer.caltech.edu/mips/psf.html; 
http://ssc.spitzer.caltech.edu/irac/psf.html} as well as Gaussian PSF models derived from measurements of the H$\alpha$ and $R$ 
images.  Since the images are all convolved with a large Gaussian kernel in order to match the 24~$\mu$m data, the resulting PSF profiles are 
nearly Gaussian, particularly in the ground-based data.  However, the aperture corrections for the ground-based and IRAC images 
are higher than they would be for unconvolved data due to the fact that our convolution does not precisely replicate the 24~$\mu$m diffraction 
lobes.  The aperture corrections were independently derived for each galaxy depending on the aperture sizes used; Table~3 
summarizes these corrections.  

Contaminant sources that show up in the 24~$\mu$m images arise from either the Rayleigh-Jeans emission of foreground stars or the 
redshifted emission of background galaxies.  We flagged objects that are located at a distance greater than $R_{25}$ from the 
center of the galaxy or that are included in the USNO catalog \citep{mon98} or 2MASS point source catalog \citep{skru06}.  
We also removed the 24~$\mu$m detections corresponding to the galaxy nucleus to avoid contamination of our region sample by embedded AGN.  
To verify the contaminant selection and ensure that the 2MASS catalog did not accidentally identify objects within the galaxy, 
we then did a visual inspection of the images to confirm or reject the flagged objects as contaminants 
based on morphology in the unconvolved IRAC and ground-based images.  

Depending on the distance of the galaxy, and therefore the luminosity sensitivity limit, the output of DAOFIND may include many 
regions that, when measured with 500~pc apertures, are significantly overlapping.  The net effect is that the same patch of a 
galaxy could be measured multiple times, and while this double-counting would not affect our results qualitatively, it would artificially 
boost the number of 500~pc regions measured in the nearest galaxies.  To limit this problem, we culled 
the region sample, weighting by a region's 24~$\mu$m flux in order to remove cases that overlapped spatially with a nearby, brighter 
region by more than 20\%.

\section{H$\alpha$ ATTENUATIONS}
Most previous measurements of dust attenuations for H~\textsc{ii} regions in
normal galaxies have been based on Balmer decrements (e.g.,
H$\alpha$/H$\beta$ ratios) derived from optical spectra \citep[e.g.,][]{mcc85, cap86, zar94}, 
or in a few instances P$\alpha$/H$\alpha$ decrements \citep{qui01, sco01}.  
However such 
studies are based on star-forming regions selected from optical
or near-infrared emission lines, and since our primary interest is
in very dusty regions, we need a method for estimating the nebular
attenuations that is valid for regions with up to several magnitudes
of visual attenuation.  A few authors have used multi-frequency radio
continuum maps together with H$\alpha$ photometry to estimate thermal 
radio fluxes and H$\alpha$ attenuations, but this method is limited 
to the brightest 10--30 H~\textsc{ii} regions in the nearest galaxies \citep[e.g.,][]{kau87, van88}.  
Instead we use the combination of mid-infrared luminosities and H$\alpha$
luminosities of the star-forming regions to derive attenuation estimates
for an infrared-selected sample.

Our method is based on the recent work by \citet{ken07a}, who
have generated a prescription for estimating H$\alpha$ attenuations 
($A_{H\alpha}$) of star-forming 
regions using measurements of the 24~$\mu$m and H$\alpha$ fluxes for individual star-forming regions.  This empirically-derived 
method treats the 24~$\mu$m flux of a region as a tracer of the obscured portion of the emission due to star formation and the 
uncorrected H$\alpha$ flux as a tracer of the unobscured portion.  The attenuation is then related to the ratio of the total 
[24$\mu$m+H$\alpha$] emission relative to the unobscured emission: 
\begin{eqnarray}
A_{H\alpha} = 2.5~\log[1+0.038~\nu L_{\nu}(24)/L(H\alpha)], 
\label{eqn:AHa}
\end{eqnarray}
where $L(H\alpha)$ stands for the uncorrected H$\alpha$ luminosity in erg s$^{-1}$ and $\nu L_{\nu}(24)$ indicates the 
``monochromatic'' 24~$\mu$m luminosity also in erg s$^{-1}$.  
The coefficient 0.038 in eq.~\ref{eqn:AHa} was calibrated using 
H$\alpha$/P$\alpha$ attenuation measurements of 42 H~\textsc{ii} regions
in M51 from \citet{cal05}.

As is emphasized in \citet{ken07a}, this method provides only an
approximate estimate of the attenuation on an object by object basis,
because the actual ratio of 24~$\mu$m luminosity to H$\alpha$
luminosity of a given object will depend on variable factors such
as the geometry of the stars, dust, and gas and the ratio of 
ionizing flux to bolometric luminosity of the OB association, which
in turn varies with the age and mass function of the embedded cluster.
This scatter can be quantified by comparing attenuations derived
from eq.~\ref{eqn:AHa} with independent measurements from the P$\alpha$/H$\alpha$
decrements or other methods.  Such comparisons have been carried
out for large samples of H~\textsc{ii} regions in M51 and M81 by \citet{cal05}
and \citet{per06} and for 220 regions in 33 
galaxies by \citet{cal07}.  Comparisons of attenuation-corrected H$\alpha$ 
fluxes with those derived via this infrared method show dispersions 
of $\pm$0.3 dex, which corresponds to uncertainties in the 
individual H$\alpha$ attenuation measurements
of $\pm$0.6$-$0.8 magnitudes.  This confirms that this method
provides only a crude estimate of the H$\alpha$ attenuation for
any individual H~\textsc{ii} region.  However since the main aim of this
paper is to identify strongly obscured regions ($A_{H\alpha} \gtrsim 3$ magnitudes),
this precision is more than adequate for our purposes.  Nevertheless
readers should bear in mind the uncertainties associated with
individual attenuation measurements.

\section{RESULTS}

\subsection{Distribution of Star-Forming Region Attenuations}

\begin{figure}
\plotone{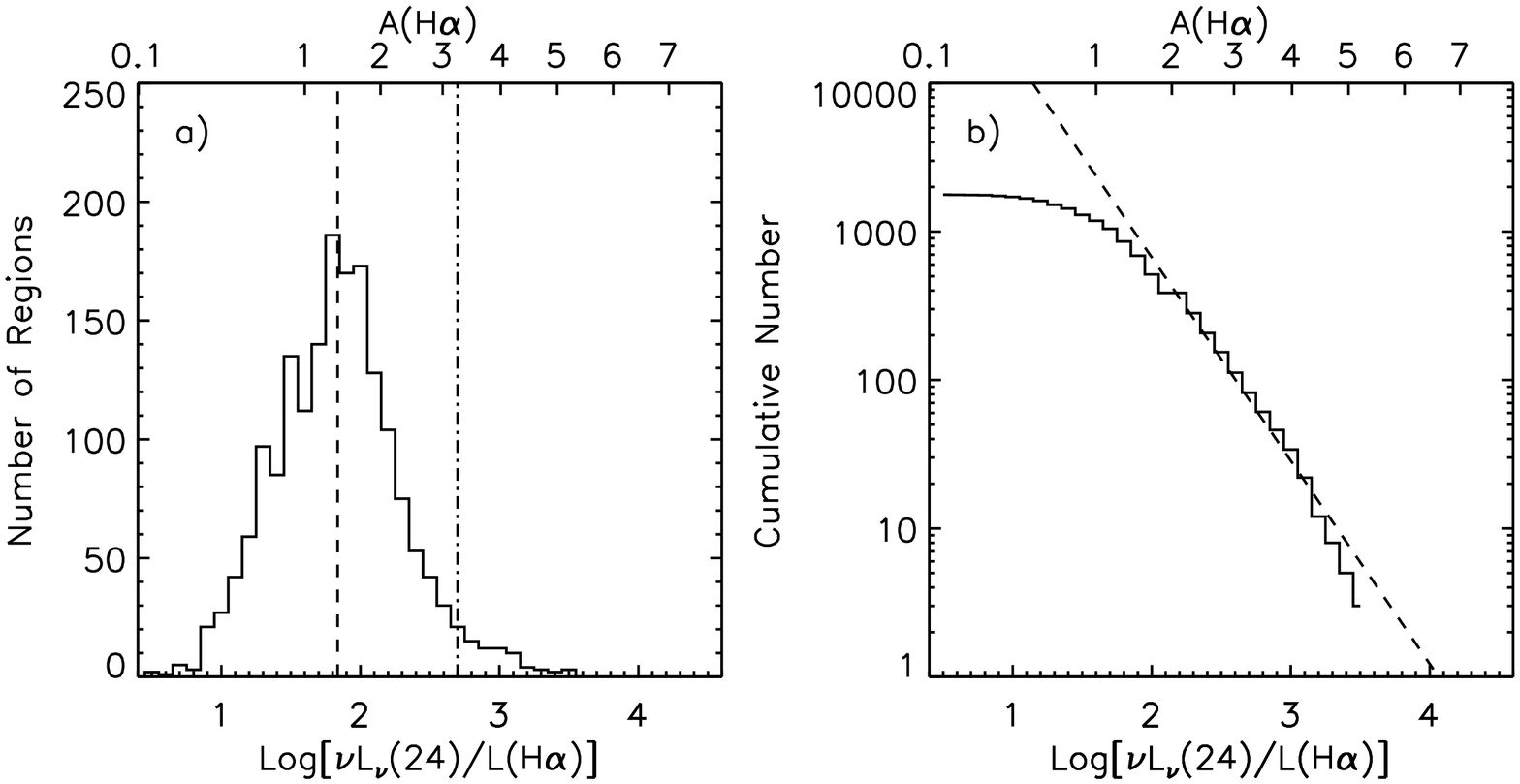}
\caption{Histograms of Log($\nu L_{\nu}(24)$/$L(H\alpha)$) or A$_{H\alpha}$ for 1772 regions selected from the sample of 
38 SINGS galaxies.  The open histogram represents the entire region sample.  The dot-dashed line is the fiducial cut proposed in the text at 
Log($\nu L_{\nu}(24)$/$L(H\alpha)$) = 2.7 (A$_{H\alpha} \sim 3.3$, A$_{V} \sim 4$), corresponding to $\sim$2 sigma above the mean of a Gaussian 
fit to the full distribution.  The panel (b) is a cumulative histogram 
of the entire region sample; the dashed line is a fit to Log($\nu L_{\nu}(24)$/$L(H\alpha)$)=[1.9,3.4] with high significance ($R^2=0.98$) yielding 
a power-law index of 1.4.  There is no large sub-population of uniformly highly obscured regions in SINGS galaxies.}  
\label{fig:ghist}
\end{figure}

In order to address the question of how many infrared-selected regions are highly obscured, we used 24~$\mu$m and H$\alpha$ photometry 
to investigate the statistics of obscured star formation on 500~pc scales in a subset of SINGS galaxies.  For our primary region sample, 
we identified a total of 1772 regions using the 24~$\mu$m datasets as discussed above.  
In Figure~\ref{fig:ghist}a we show a histogram of Log[$\nu L_{\nu}(24)$/L(H$\alpha$)] 
along with the corresponding H$\alpha$ attenuations.  From the distribution of attenuations it is 
immediately clear that we do not see a large sub-population of highly-obscured star-forming regions within this infrared-selected sample 
on 500~pc spatial scales.  The median value is 1.8 which corresponds to an H$\alpha$ attenuation of 1.4 magnitudes 
(A$_{V} \sim$ 1.7), and there is a slight tail to attenuations higher than 3-4 magnitudes (skew = 0.35).  
Contaminant sources such as foreground stars or background galaxies would also populate the high attenuation portion of parameter space; 
to address this concern, we carefully verified that contaminant 
sources had been removed from the sample using the approach described in Section 4.  
At this point, we must make clear what is meant by the ``highly obscured
population" or ``subpopulation."  Our beam encompasses many tens or hundreds
of massive stars (see Table~1, minimum luminosities) and entire giant molecular clouds (GMCs).  
If all the stars are uniformly attenuated, then $A_{H\alpha}$ in Figure~\ref{fig:ghist} represents
the attenuation to each star.  However, if, as is likely, some of the stars
lie outside the GMC (due to breakout, blister H\textsc{ii} regions, etc.) and
some of the stars are embedded, then the $A_{H\alpha}$ represents an average
attenuation, weighted heavily toward the unattenuated stars.  
In this study, we do not find evidence for any large subpopulation of 
uniformly highly-obscured star-forming regions on 500~pc scales; we discuss 
the question of smaller scales in Section 7.1.2.  

Figure~\ref{fig:ghist}b presents the same data in the form of a cumulative histogram.  
For illustration, the cumulative distribution can be fit by a simple power law with an index of $\sim$1.4 with high 
significance ($R^2=0.98$) suggesting that the higher attenuation regions are simply an extension of a smooth distribution.  
To quantify the fraction of regions in the highly-obscured tail relative to the entire sample, we make a 
cut at Log[$\nu L_{\nu}(24)$/L(H$\alpha$)]$ = 2.7$ (A$_{H\alpha} \sim$ 3.3, A$_{V} \sim 4.0$).  This cut corresponds to $\sim$2 times the 
standard deviation of a Gaussian fit to the overall region distribution.  By this definition, the number of regions in the sample 
found in this highly-obscured tail of the distribution is 69, $\sim$4\% of the sample.  

\begin{figure}
\plottwo{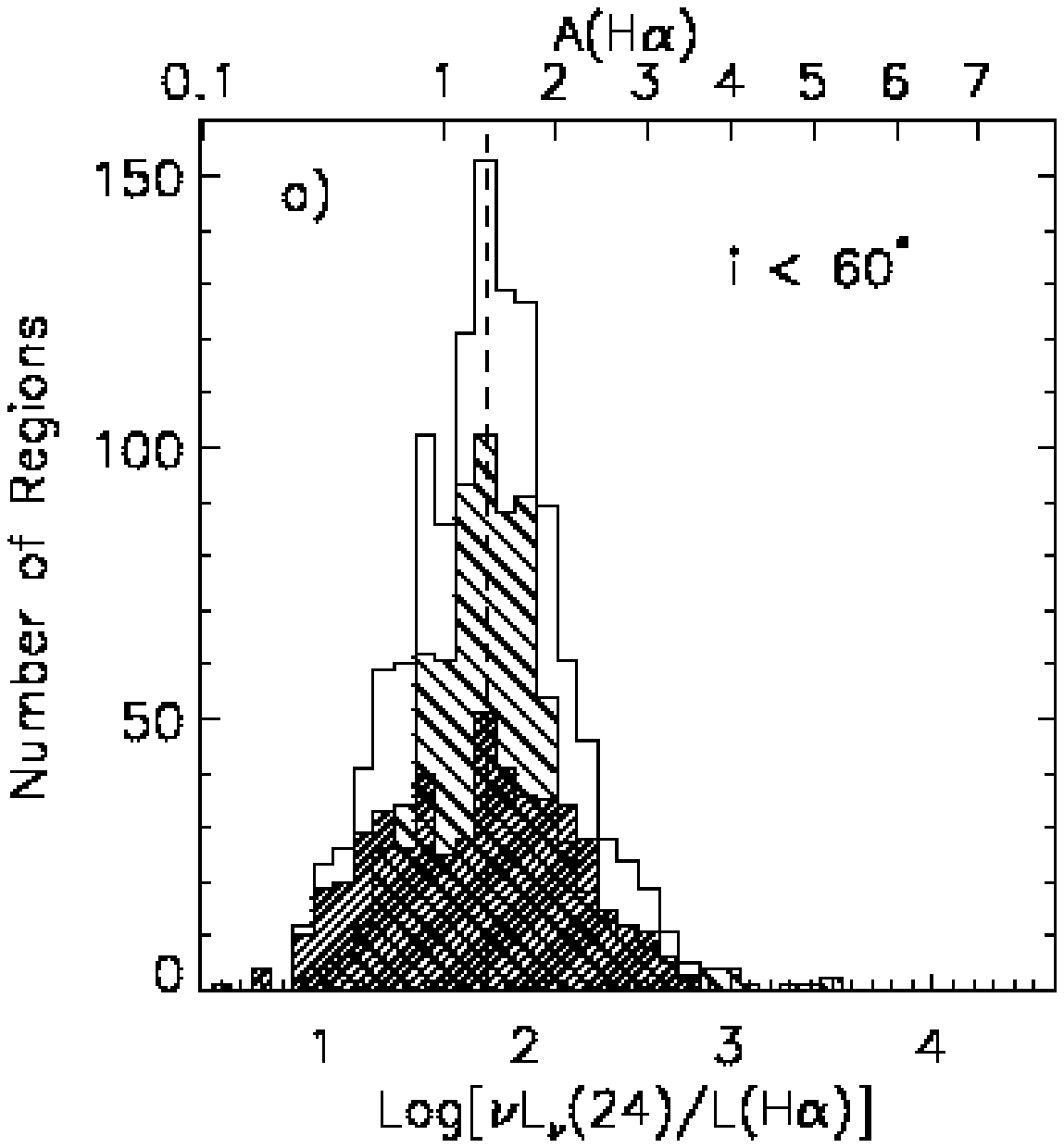}{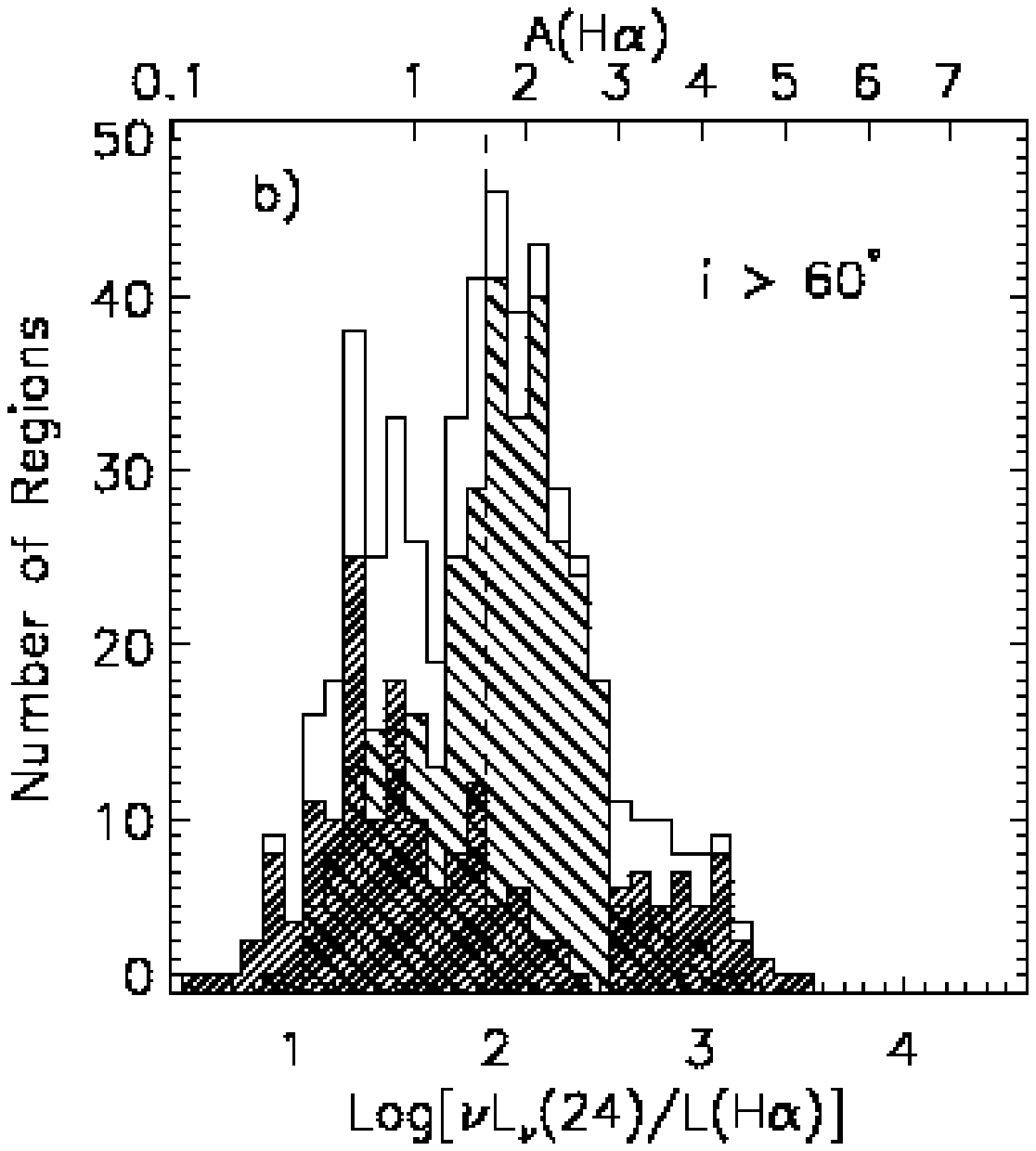}
\caption{Attenuation distributions (Log($\nu L_{\nu}(24)$/$L(H\alpha)$) or A$_{H\alpha}$) are shown separated by host galaxy 24~$\mu$m 
luminosity for the low inclination galaxies ($i > 60^{\circ}$; (a)) 
and high inclination galaxies ($i > 60^{\circ}$; (b)).  In each panel the narrow-hatched histogram is the low luminosity bin 
($\nu L_{\nu}(24) < 8.4\times10^{42}$ [erg s$^{-1}$]), while the wide-hatched histogram the high luminosity bin ($\nu L_{\nu}(24) > 8.4\times10^{42}$ [erg s$^{-1}$]), 
and the open histogram includes all luminosities.  The region attenuation distribution is more strongly peaked for low inclination 
galaxies, while high inclination cases populate the wings of the distribution.  The dashed line in each panel represents the median value of the distribution.  
Highly-obscured regions are found in both high and low luminosity galaxies.}
\label{fig:gcolorlumgal}
\end{figure}

Thus far, we have treated the entire sample of regions as one; 
however, the attenuation distribution changes slightly if we divide the sample by 
inclination and luminosity.  In Figure~\ref{fig:gcolorlumgal}, we have split the sample into two panels 
by the inclination of the host galaxy, where high inclination is defined here as $i > 60^{\circ}$.  The 
median Log[$\nu L_{\nu}(24)$/L(H$\alpha$)] values from the low and high inclination subsets are 
essentially equal within the errors --- 1.8 for the low inclination and 1.9 for the high inclination cases.  
More striking, however, is the difference in shape between the two distributions: the low inclination 
subsample distribution is more sharply peaked (kurtosis=0.71, with 0 signifying a normal distribution) than that of 
the high inclination subsample (kurtosis=-0.28).  This likely reflects the fact that the use of 
our empirical attenuation method becomes more complicated at high inclination; measured attenuations 
may be boosted in some cases due to increased pathlength through the disk, causing larger scatter.  
In addition, at the highest inclinations the region selection will be affected by confusion, 
and there may be a bias towards selecting more regions from the near side of the galaxy.  
In each panel of Figure~\ref{fig:gcolorlumgal}, we divide the subsample into two bins of total galaxy 24~$\mu$m luminosity, 
divided at the region-weighted mean galaxy 24~$\mu$m luminosity of the sample (8.4$\times$10$^{42}$ erg~s$^{-1}$).  
The region attenuation distributions are similar to that of the full sample, with minimal change in the 
median attenuation as a function of galaxy luminosity.  Thus, highly-obscured regions are drawn 
from both high and low luminosity galaxies.

\begin{figure}
\plotone{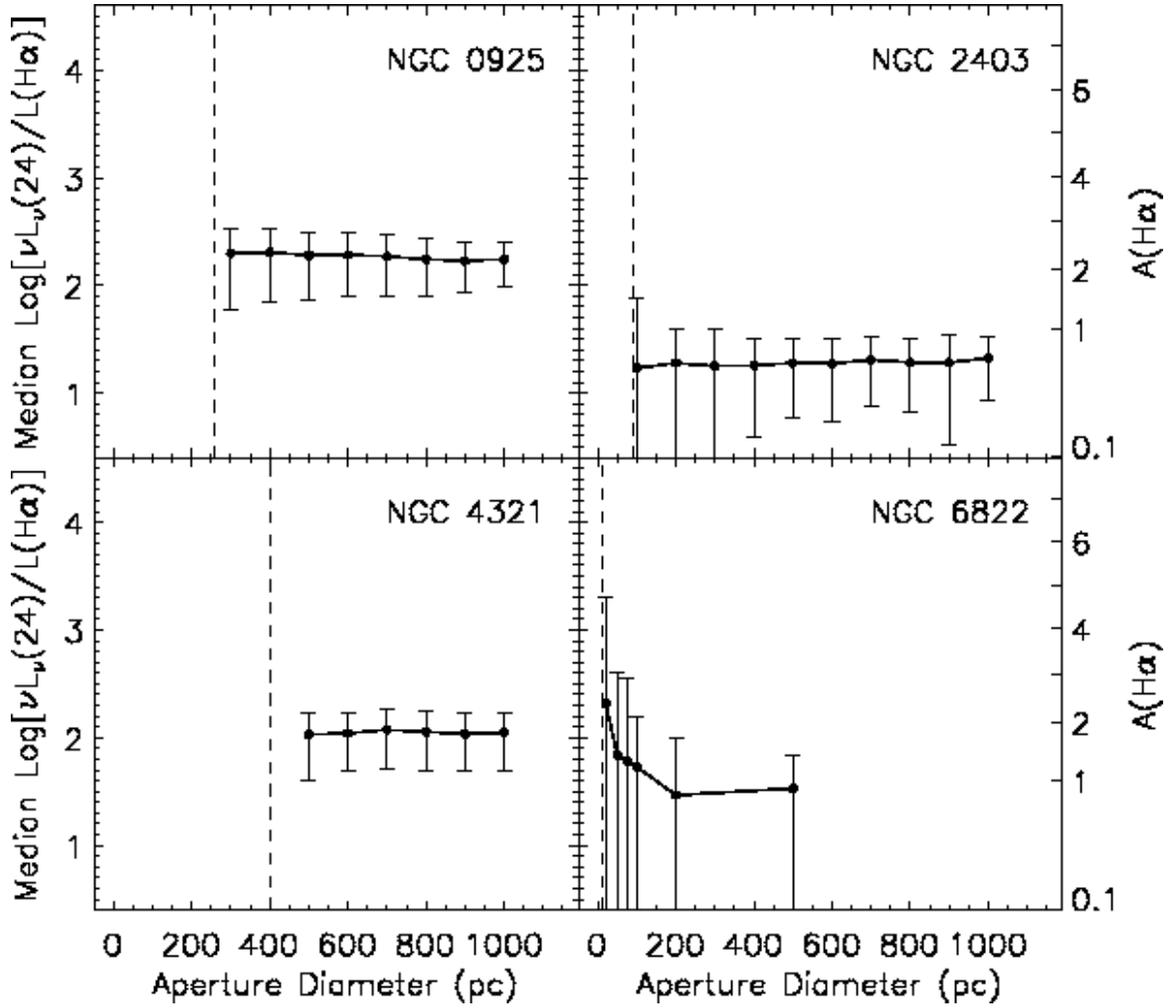}
\caption{Median attenuation for a representative set of galaxies as a function of aperture diameter used.  
The error bars indicate the standard deviation of all the region measurements included in the median.  The dashed line 
indicates the size of the FWHM of the 24~$\mu$m PSF, below which the measurement is unresolved.  The median attenuation is 
nearly constant with aperture size between 100 and 1000~pc scales.  In the sole case where we can probe below 100~pc, the dwarf galaxy 
NGC~6822, the measured attenuation appears to climb substantially; however, at these scales we are affected by the individual region morphology.  
For large apertures ($>$100~pc) the measured value of Log($\nu L_{\nu}(24)$/$L($H$\alpha)$) is relatively insensitive to aperture size.}  
\label{fig:gapsize}
\end{figure}

In this study we are interested in the attenuation properties of star-forming regions on large, 500~pc scales.  
It is important to note, however, that this aperture may be much larger than the size of a given star-forming region as viewed 
at 24~$\mu$m.  
To get a sense of how the chosen aperture size affects our results, we have done photometry at scales spanning between roughly 200 and 
1000~pc for most of the sample, between 100 and 1000~pc for NGC~2403 (a particularly nearby case at 3.2 Mpc), and down to 20~pc 
for NGC~6822 (a dwarf galaxy at 0.5 Mpc).  In Figure~\ref{fig:gapsize} we plot the median galaxy 24$\mu$m/H$\alpha$ ratio versus aperture size for 
a representative set of 4 galaxies: NGC~0925, NGC~2403, NGC~4725, and NGC~6822.  
The vertical dashed line represents the point at which the aperture diameter is equal to the FWHM of the 24~$\mu$m PSF, and below which 
the aperture measurement effectively undersamples the PSF.  We see very little variation in the 
median attenuation 
down to 100~pc scales, while the scatter increases slightly over this range with decreasing aperture size.  In our most nearby case, 
NGC~6822, we do see a 
significant rise in our measured attenuation as we probe below 100~pc, but at these scales we are also likely seeing the 
effects of the complex morphologies of H$\alpha$ structures - shells and bubbles - that may be offset spatially from the brightest 
corresponding 24~$\mu$m source.  For the purposes of the current study, this analysis illustrates the fact that our application of 
the empirical method applies to large scales, i.e. greater than 100~pc, where the results are fairly insensitive to the 
particular aperture diameter chosen.  

\subsection{Infrared Colors of Star-Forming Regions}

\begin{figure}
\plotone{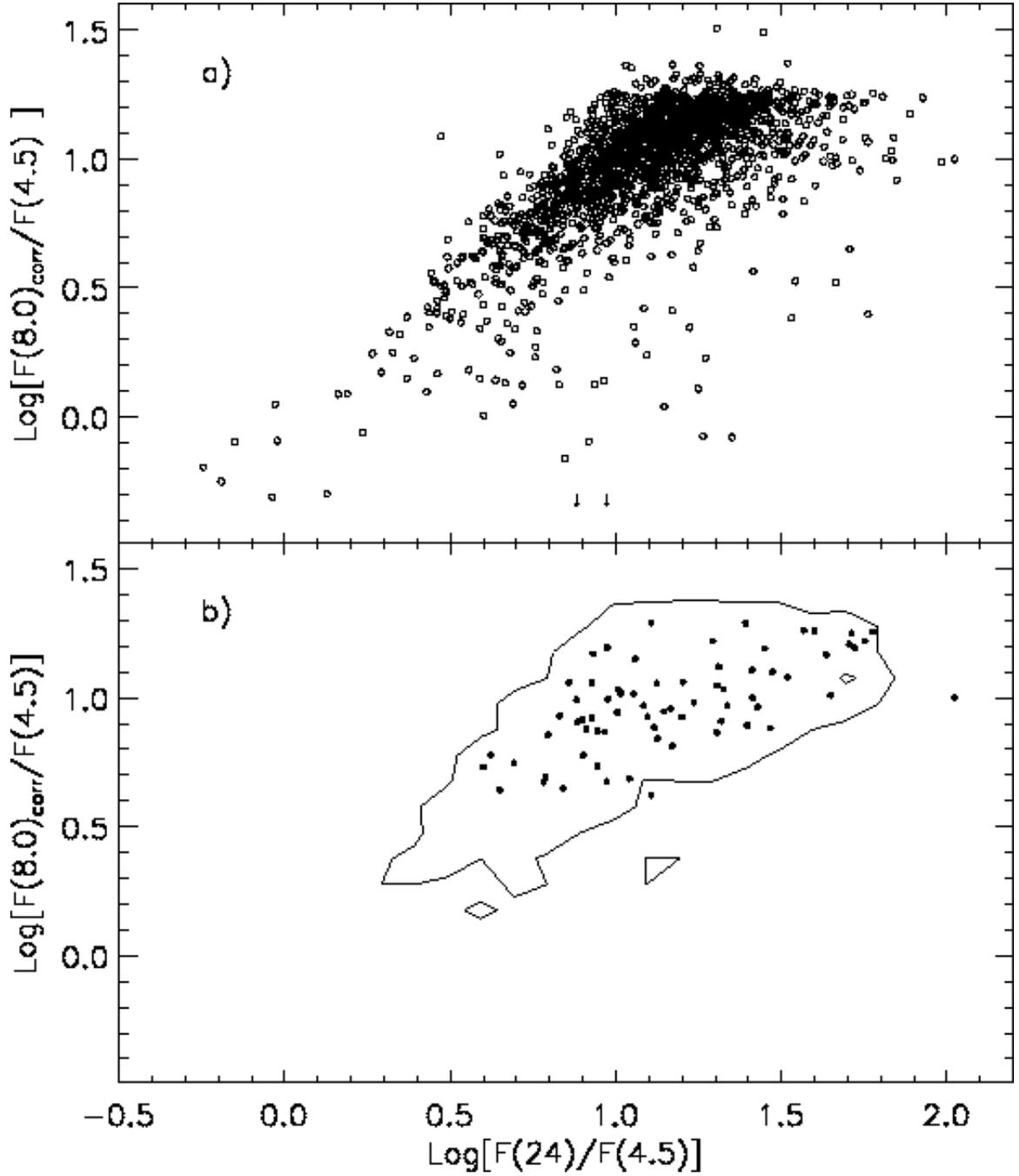}
\caption{(a)  F(8.0)$_{corr}$/F(4.5) versus F(24)/F(4.5) plot for the entire region sample.  (b)  F(8.0)$_{corr}$/F(4.5) 
versus F(24)/F(4.5) plot for the 
highly-obscured subset, as defined by the attenuation cut described in the text.  The contour encloses 95\% of the total region sample.  
F(8.0)$_{corr}$ values have been corrected for the contribution of stellar emission, as discussed in the text.  
Flux densities are F$_{\nu}$ quantities in erg s$^{-1}$ Hz$^{-1}$.  Highly-obscured regions do not stand out from the rest of the region 
sample in terms of their infrared colors.}  
\label{fig:gcolors}
\end{figure}

The infrared colors of the entire sample of regions are presented as flux density ratios in Figure~\ref{fig:gcolors}.  
Following the discussion of \citet{dop05} we have chosen 
F(8.0)/F(4.5) versus F(24)/F(4.5) as a representative pair of colors that include contributions from the 
stellar (4.5~$\mu$m), PAH (8.0~$\mu$m), and dust (24~$\mu$m) components of a given region, 
although we note that the IRAC bands do not sample the 10~$\mu$m silicate feature.  
We have corrected the 8.0~$\mu$m flux densities for the contribution of stellar emission using the following relation: 
\begin{equation}
F(8.0)_{corr} = F(8.0) - 0.232 \times F(3.6)
\end{equation}
where F(8.0)$_{corr}$ is the corrected 8.0~$\mu$m flux density, F(8.0) is the observed value, 
and F(3.6) is the 3.6~$\mu$m stellar flux density \citep{hel04}.  
The contribution of dust emission to the 3.6 and 4.5~$\mu$m bands is assumed to be negligible 
for the purposes of this work.  We have not applied any additional correction for the complex scattering 
that occurs when light from extended sources is incident on the IRAC focal 
plane\footnote{http://ssc.spitzer.caltech.edu/irac/calib/extcal/} and that cause flux measurements to be systematically overestimated.  
The effect is negligible for the more distant galaxies in the sample, but even for the nearby cases 
the magnitude of the shift is less than 0.11 dex in Log[F(8.0)/F(4.5)] and less than 0.03 dex in Log[F(24)/F(4.5)]; 
this offset does not affect our conclusions. 
The contour encloses 95\% of the total sample of regions, and the open and filled circles represent the 
individual regions and the highly-obscured subset, respectively.  
We can see that the highly-obscured regions do not stand out from the rest of the region sample in terms of their 
infrared colors.  
The positive correlation seen between the F(8.0)/F(4.5) and 
F(24)/F(4.5) appears to plateau at Log[F(8.0)/F(4.5)]$\sim$1.25.  This behavior is seen in the region sample as a whole as 
well as in some of the individual galaxies.  
One plausible explanation is that at low fluxes, both the 8~$\mu$m PAH emission and the 24~$\mu$m emission are 
excited by single optical/UV photons and are therefore proportional to the number 
of absorbed stellar photons.  This would imply that at low values of $F(24)$ one would 
expect a linear correlation of $F(8.0)$ to $F(24)$, as is observed.  
As the 24~$\mu$m emission increases, however, larger grains could be
sufficiently heated to provide a contribution to 24~$\mu$m emission from
the Wien tail of this steady state warm dust.  This then, may provide
the plateau (increasing 24~$\mu$m emission versus constant 8~$\mu$m emission).

\begin{figure}
\plotone{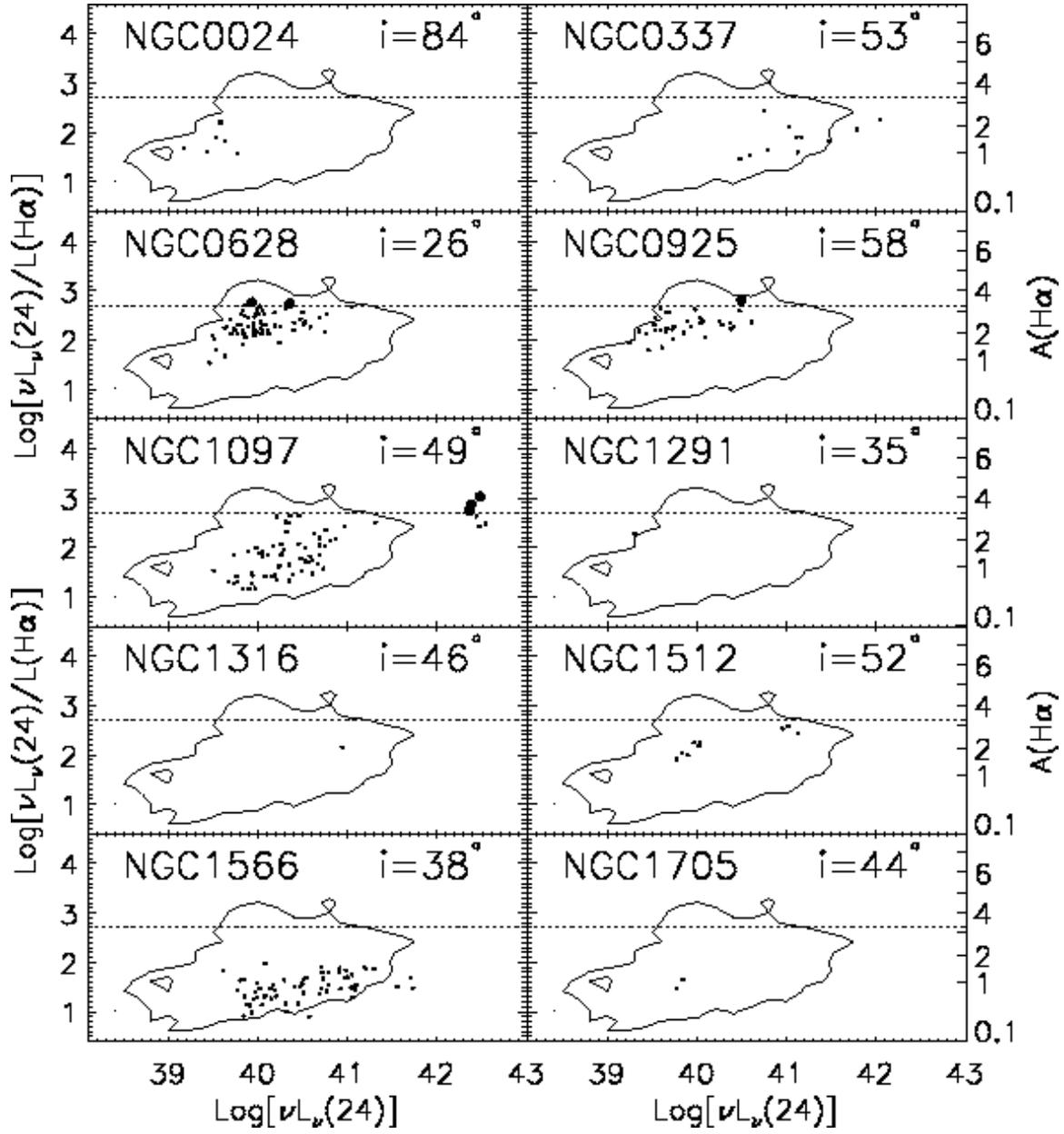}
\caption{Log[$\nu L_{\nu}(24)$/$L($H$\alpha)$] versus Log[$\nu L_{\nu}(24)$] for the region sample, divided by host galaxy.  
The contour is common to all panels and encloses 95\% of the total region sample, while 
the small points are regions within the individual galaxy.  The dotted line is the attenuation cut defined in the text, and 
the filled circles represent those regions in the galaxy that have been 
classified as highly obscured.  The galaxy inclination is given in each panel.  Highly attenuated regions tend to be more luminous.}  
\label{fig:gcolorsflux}
\end{figure}

\begin{figure}
\plotone{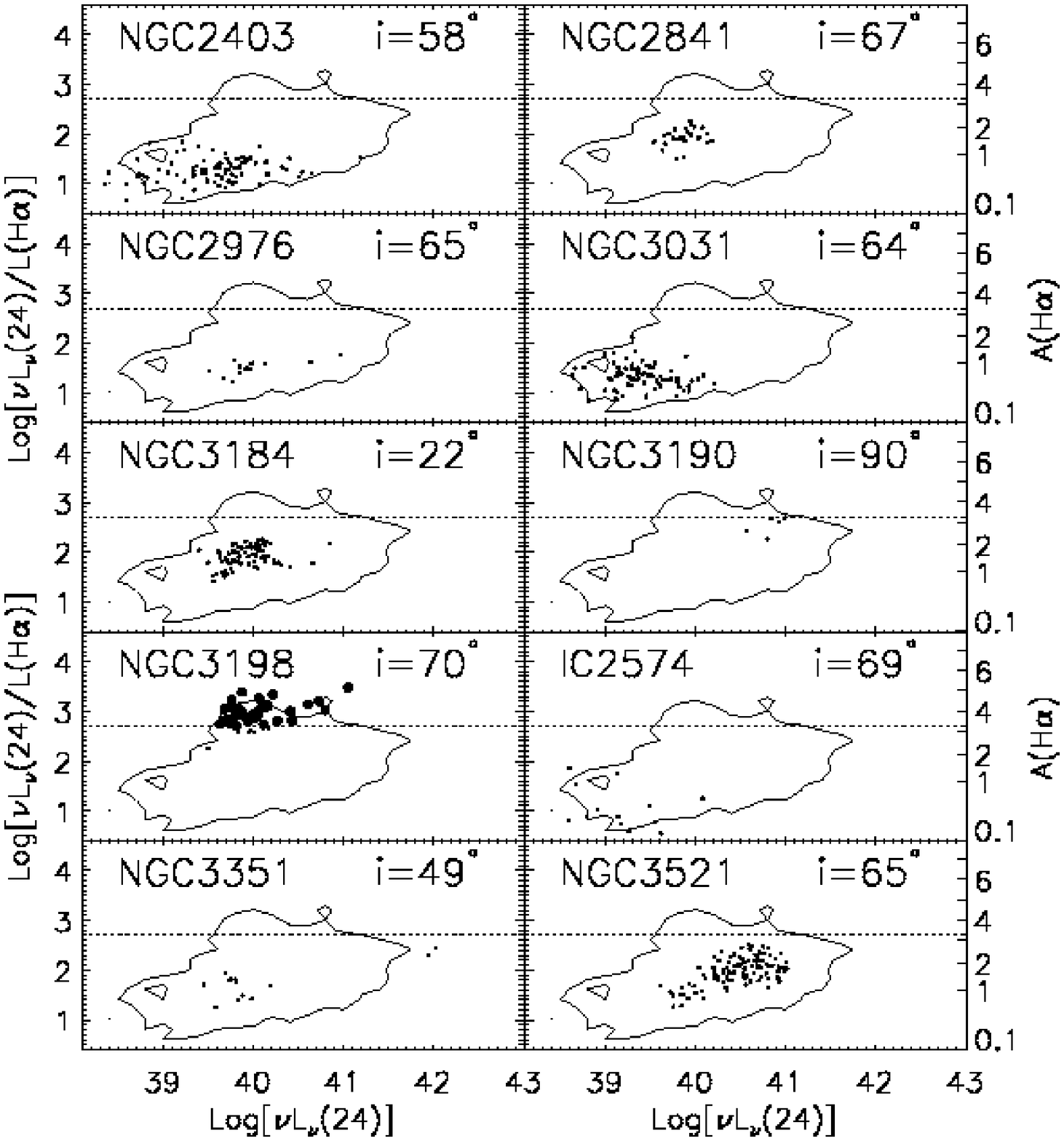}
\caption{Same as Figure~\ref{fig:gcolorsflux}.}
\label{fig:gcolorsflux2}
\end{figure}

\begin{figure}
\plotone{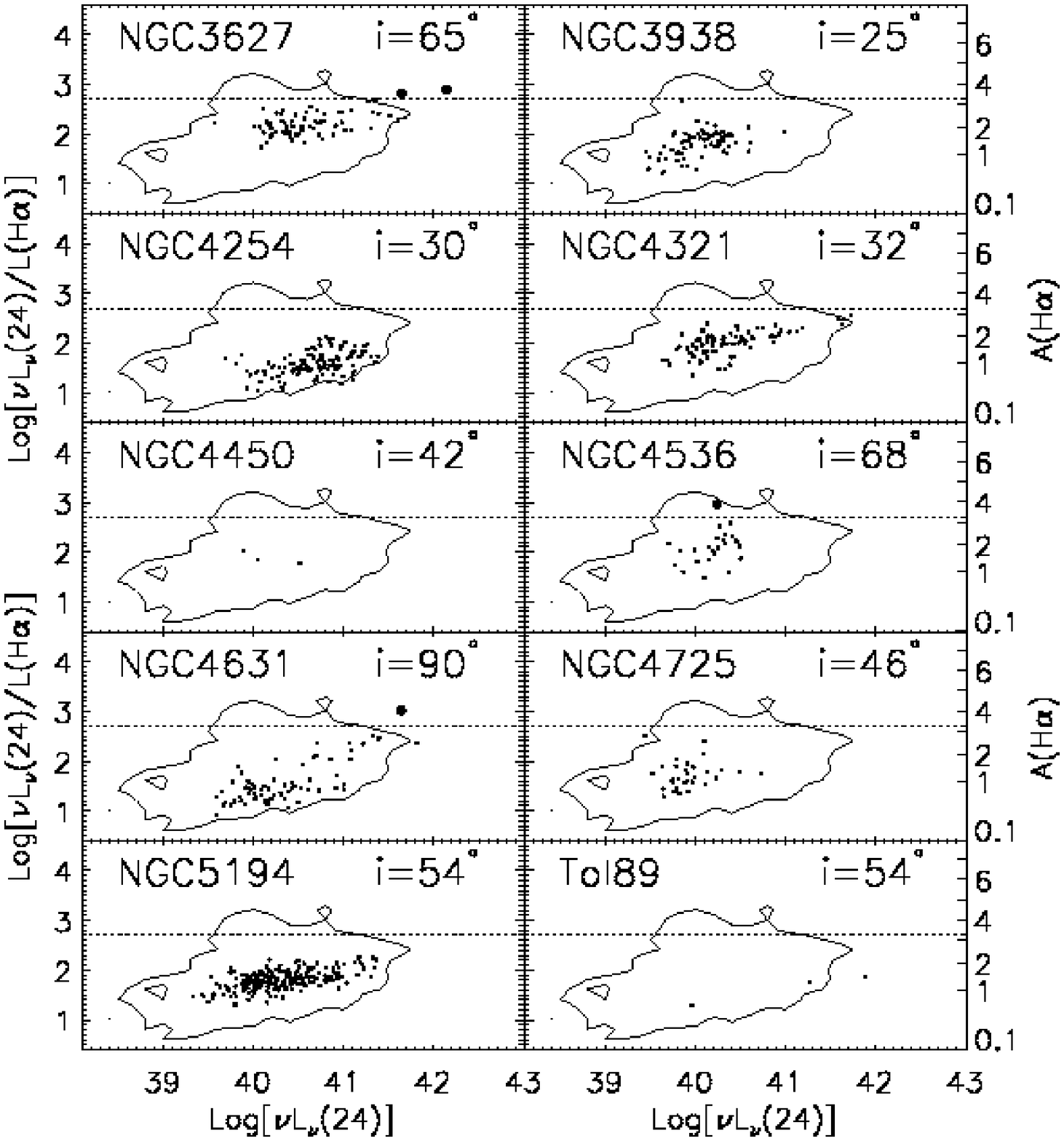}
\caption{Same as Figure~\ref{fig:gcolorsflux}.}
\label{fig:gcolorsflux3}
\end{figure}

\begin{figure}
\plotone{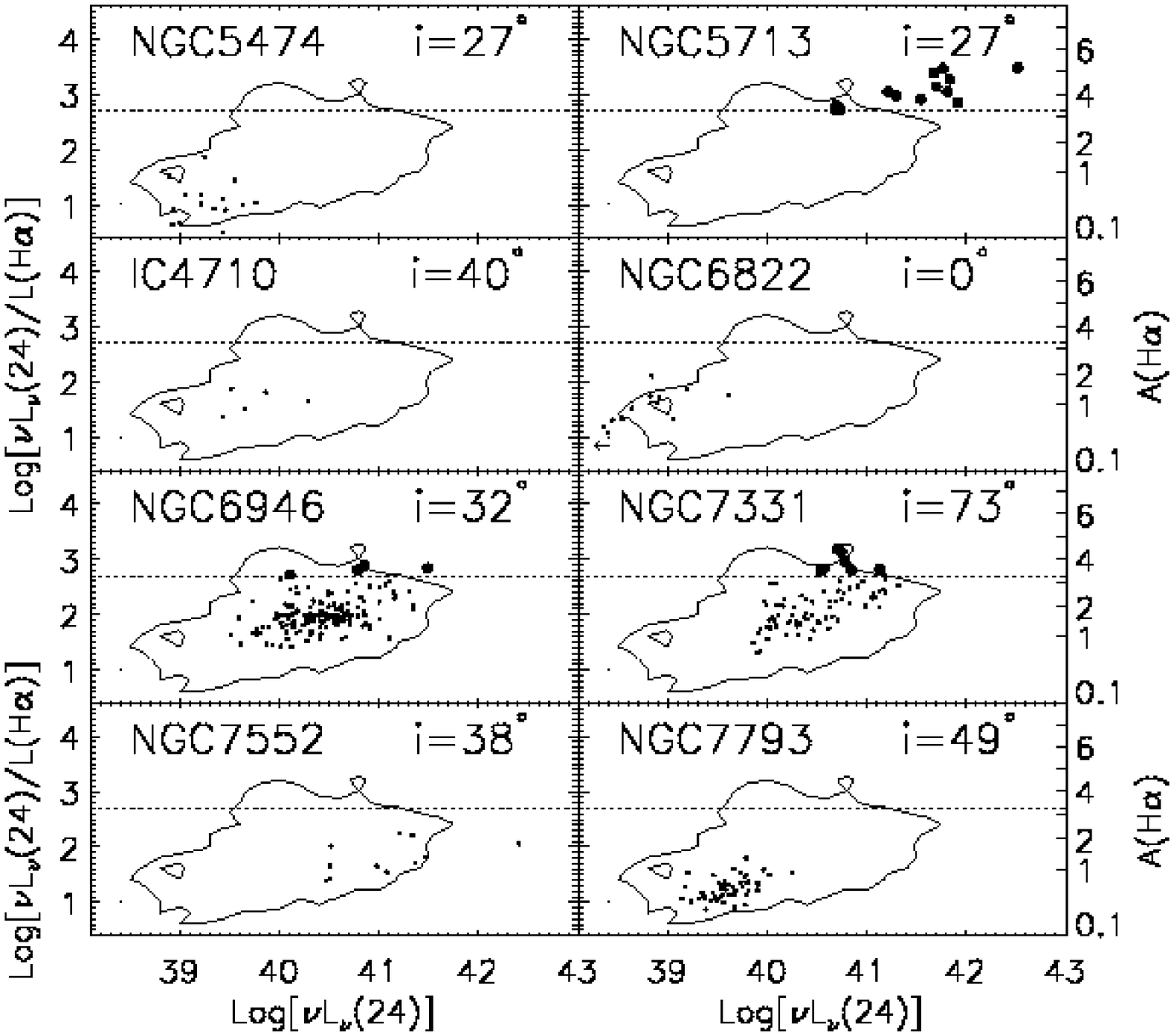}
\caption{Same as Figure~\ref{fig:gcolorsflux}.}
\label{fig:gcolorsflux4}
\end{figure}

In Figures~\ref{fig:gcolorsflux}-\ref{fig:gcolorsflux4}, we plot the measured 24$\mu$m/H$\alpha$ attenuation versus the region 24~$\mu$m luminosity, again 
using a 95\% contour to represent the entire region sample; the dotted line represents the subsample defined by attenuation in 
the previous section (A$_{H\alpha} > 3.3, A_{V} > 4.0$).  
In the entire sample, we see a general trend for more attenuated regions to be more luminous, a trend that is roughly reproduced within 
individual galaxies, with a few exceptions like NGC~3031.

\subsection{Properties of the Most Obscured Regions}

In applying the arbitrary cut from Section 6.1 (A$_{H\alpha} > 3.3, A_{V} > 4.0$), we were aiming to illustrate the relative 
paucity of uniformly highly obscured regions in the sample ($\sim$4\%), 
but we did not mean to imply that they are physically distinct from the rest of 
the population.  To demonstrate this fact we have further characterized the regions that make up the tail of the attenuation distribution 
(see Figure~\ref{fig:ghist}).  In Table~1, we list for each galaxy the total number of regions selected, the 24~$\mu$m luminosity of the faintest region measured, 
and the number of highly-obscured regions using our arbitrary definition.  The number of such regions in a galaxy is typically small, and 
there are many galaxies with none.  The exceptions are NGC~5713 and NGC~3198, in which all but a handful of the measured regions qualify as 
highly obscured by our definition.  As discussed in the next section, these two cases show globally high attenuation relative to the rest of 
the galaxy sample.   

\begin{figure}
\plotone{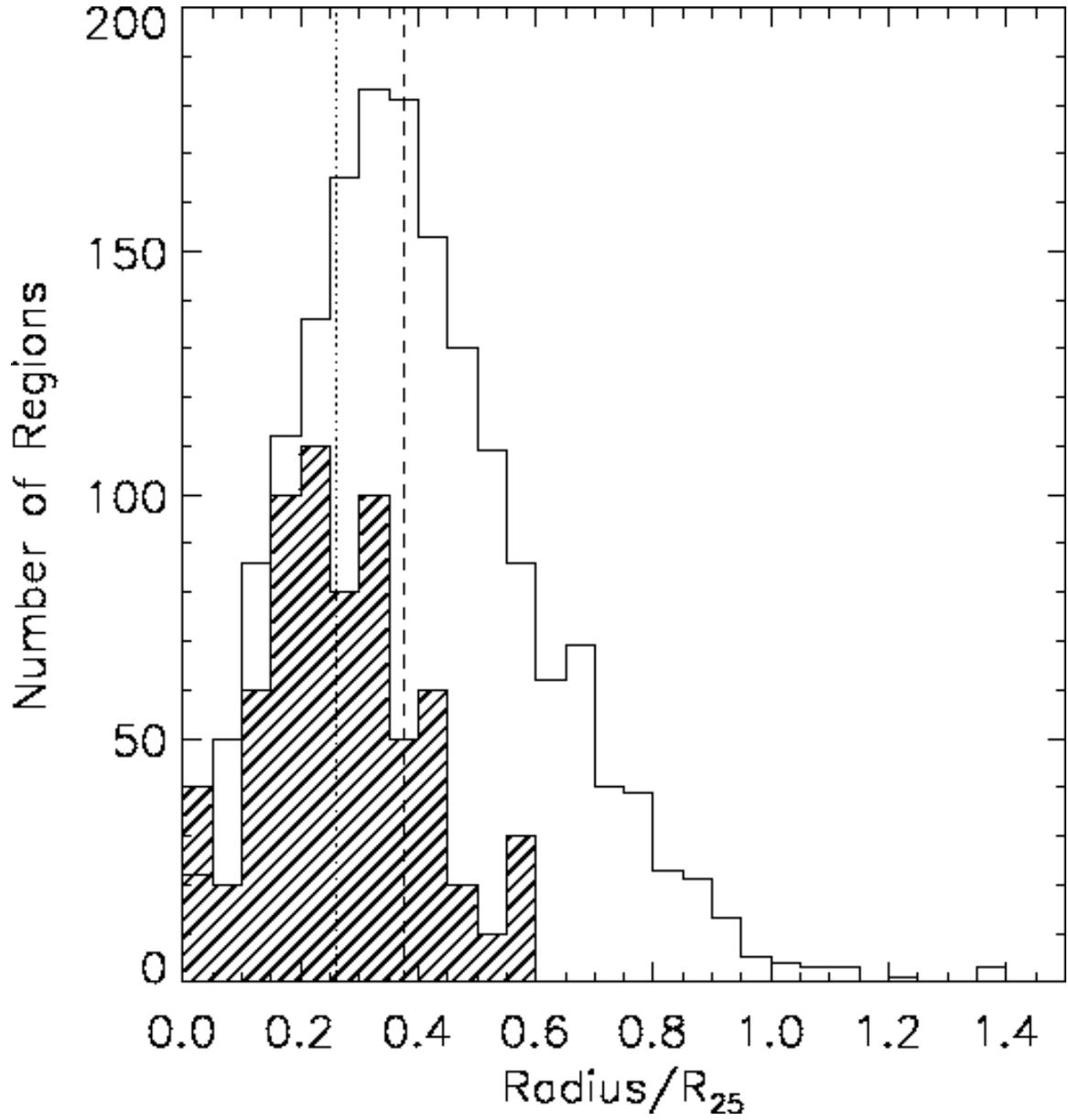}
\caption{Histogram of highly-obscured region locations in the host galaxy, 
given as a deprojected radial position in units of R$_{25}$ (hatched histogram; multiplied by a factor of 10 for clarity).  
The open histogram shows the corresponding locations for the entire region sample.  
The dashed line shows the median location for the entire sample (0.38 R$_{25}$) while the dotted line is the median location for more 
highly-obscured regions (0.26 R$_{25}$).  More highly-obscured regions are generally located closer to the center of the host galaxy.}  
\label{fig:ghistlocat}
\end{figure}

The first question we ask is where the more highly-obscured regions are located within the host galaxy.  In Figure~\ref{fig:ghistlocat} 
we present a histogram of the deprojected radius of the more highly-obscured regions within the host galaxy compared with 
the entire region sample (excluding the edge-on cases NGC~4631 and NGC~3190).  The histogram for the 
full sample extends out to 1.38 $R_{25}$ and has a median radius around 0.38 $R_{25}$.  
In contrast, the population in the highly-obscured tail is only found within 0.59 $R_{25}$ with a median position of 0.26 R$_{25}$.  
Thus, the highly-obscured subset is drawn primarily from the inner parts of the host galaxies.   
To characterize these regions further, we investigate the colors of the highly-obscured tail compared with those of the rest 
of the regions in the host galaxy and with those of the entire region sample.  Returning to Figure~\ref{fig:gcolors}, we see that for the most part, the 
highly-obscured cases do not stand out in terms of infrared colors, but instead are scattered throughout the rest of the region color 
distribution of the host galaxy.  In Figures~\ref{fig:gcolorsflux}-\ref{fig:gcolorsflux4}, we see that the loose trend between region luminosity and attenuation 
means that the regions which make up the tail of the attenuation distribution tend to also be more luminous.  These comparisons all suggest 
that the more highly obscured regions in our sample are not a distinct sub-population; rather, they are simply the bright, central 
regions of their host galaxies.

\subsection{Radial Attenuation Gradients}

\begin{figure}
\plotone{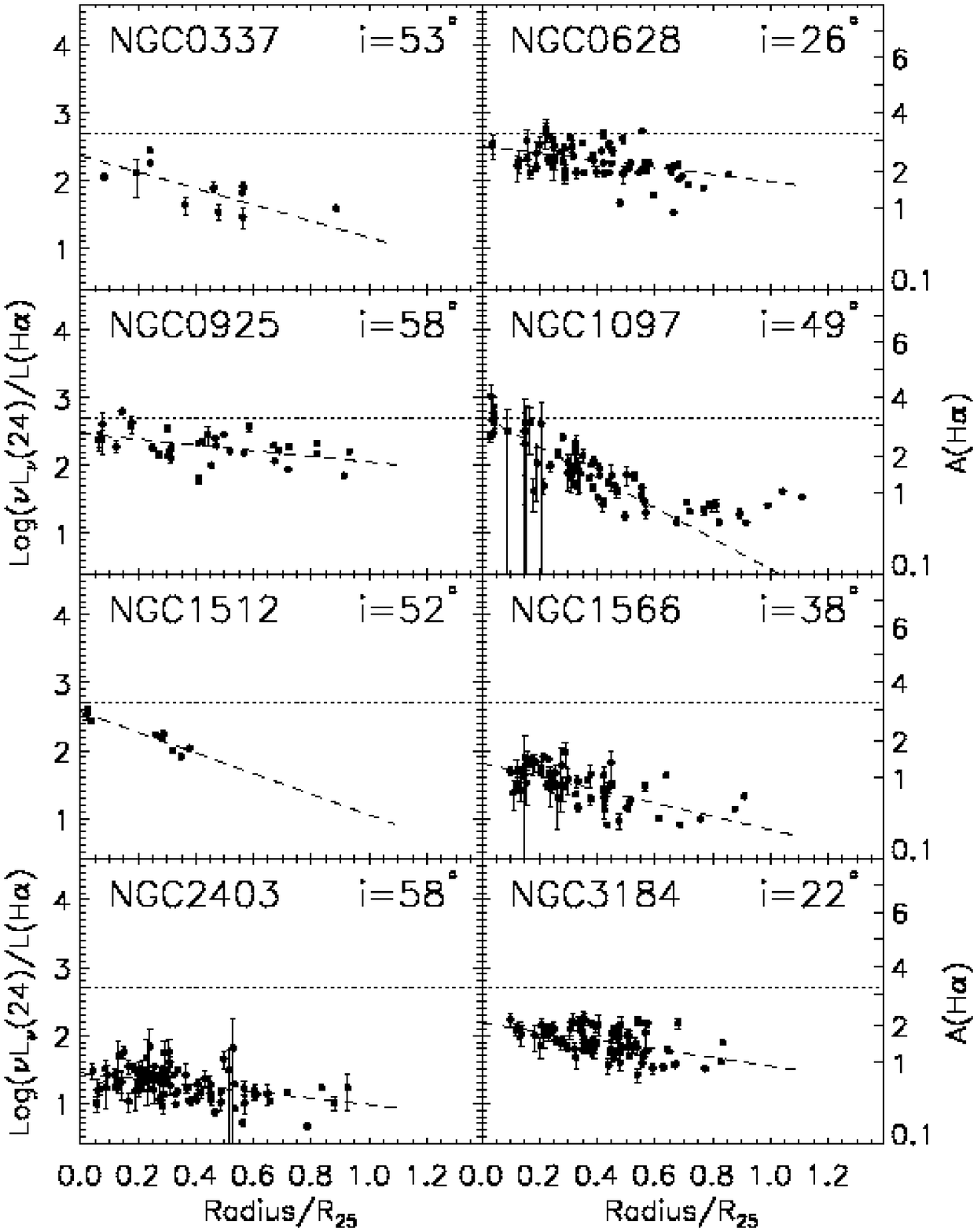}
\caption{Deprojected radial profiles of Log($\nu L_{\nu}(24)/L($H$\alpha)$) for galaxies in the sample with inclinations less than 60$^{\circ}$.  In cases 
where the region profile can be well fit by linear regression, the dashed line gives the linear fit to the radial range 0.0--0.65 R$_{25}$.  
The dotted line corresponds to the attenuation cut (A$_{H\alpha}>$3.3) applied in Section 6.1.}  
\label{fig:iradialfitlow}
\end{figure}

\begin{figure}
\plotone{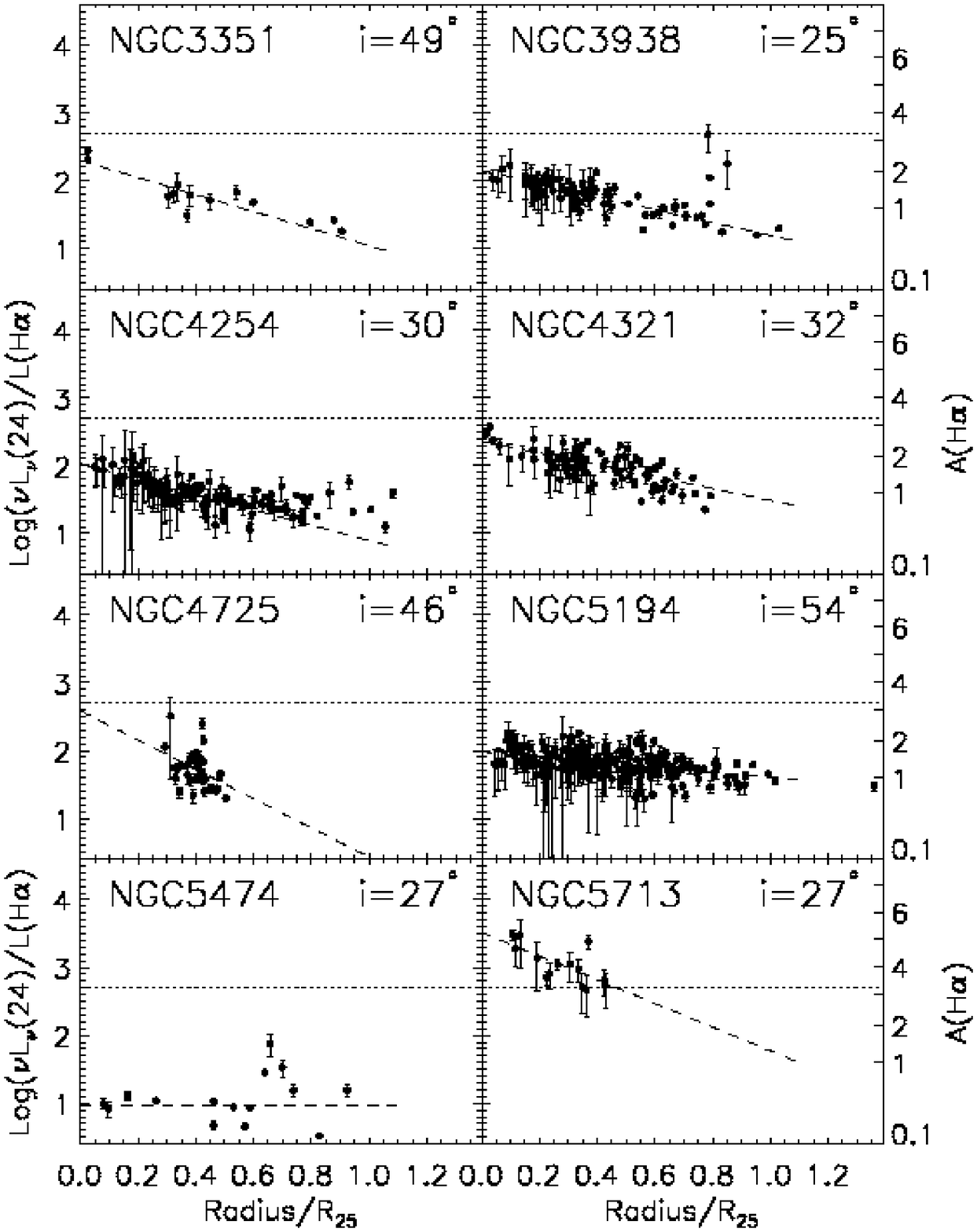}
\caption{Same as Figure~\ref{fig:iradialfitlow}.}
\label{fig:iradialfitlow2}
\end{figure}

\begin{figure}
\plotone{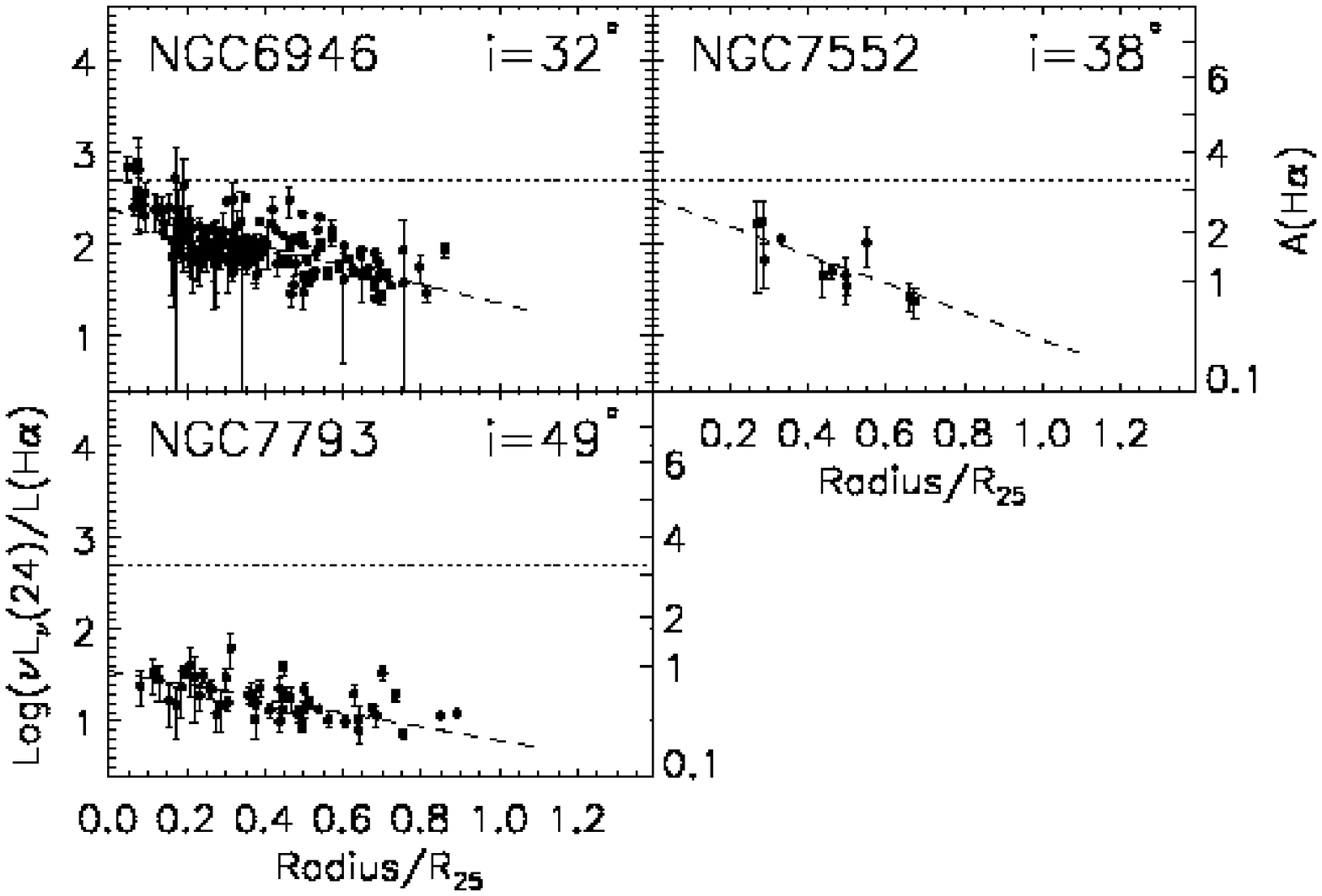}
\caption{Same as Figure~\ref{fig:iradialfitlow}.}
\label{fig:iradialfitlow3}
\end{figure}

\begin{figure}
\plotone{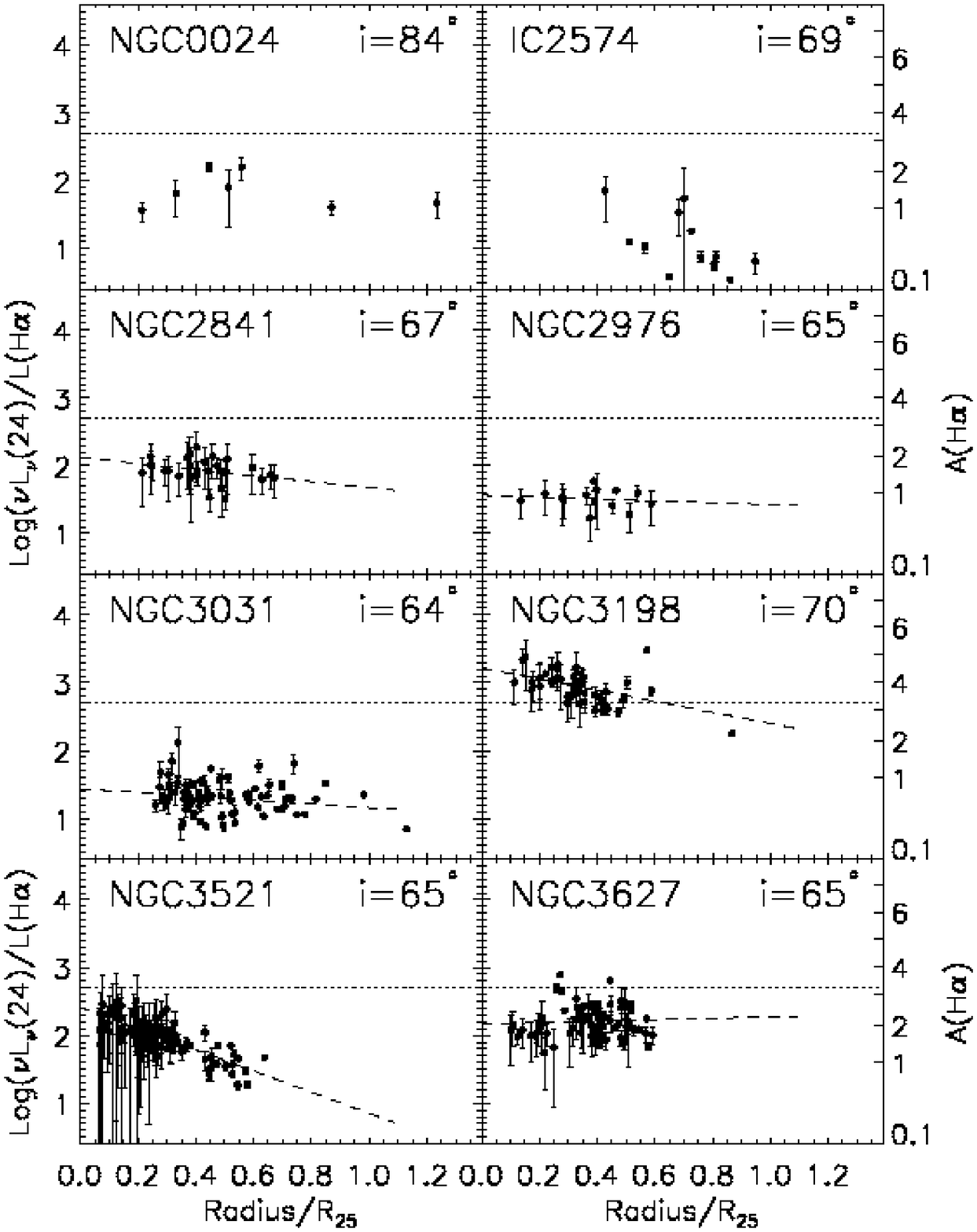}
\caption{Deprojected radial profiles of Log($\nu L_{\nu}(24)/L($H$\alpha)$) for galaxies in the sample with inclinations greater than 60$^{\circ}$.  In cases 
where the region profile can be well fit by linear regression, the dashed line gives the linear fit to the radial range 0.0--0.65 R$_{25}$.  
The dotted line corresponds to the attenuation cut (A$_{H\alpha}>$3.3) applied in Section 6.1.}  
\label{fig:iradialfithigh}
\end{figure}

\begin{figure}
\plotone{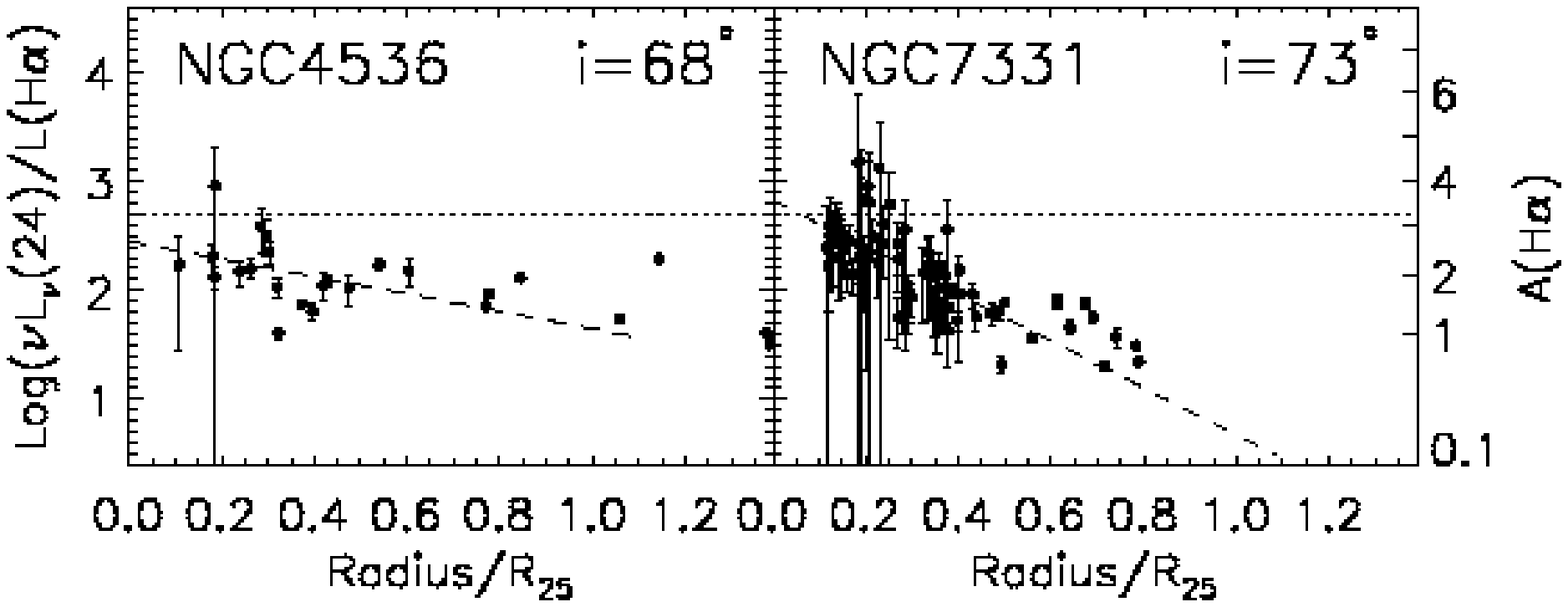}
\caption{Same as Figure~\ref{fig:iradialfithigh}.}
\label{fig:iradialfithigh2}
\end{figure}

Looking at the global properties of our galaxy sample, we find that a majority of the galaxies show a distinct and nearly 
linear decreasing radial trend in 24$\mu$m/H$\alpha$, a finding consistent with other studies \citep{boi04, boi05, pop05, rou05}.  
Figures~\ref{fig:iradialfitlow}-\ref{fig:iradialfithigh2} illustrate this result by showing the 24$\mu$m/H$\alpha$ ratio versus deprojected radius for each galaxy, with the 
galaxy sample split into two groups by inclination.  
The error bars are the standard errors of the mean derived using the three different background measurements described in Section 4.  
Galaxies that do not contain enough regions to determine the extent of a radial trend 
are not shown (NGC~1291, NGC~1316, NGC~1705, NGC~4450, Tololo~89, IC~4710); NGC~6822 is similarly excluded due to complicated morphology.  
The edge-on nature of NGC~4631 and NGC~3190 preclude an accurate radial deprojection.
In cases where the region profile is well-described by linear regression we have shown the best linear fit to the radial range 
0.0--0.65 $R_{25}$.  
In a few cases, the region profile is not adequately described by a linear fit due 
to more complex structure (e.g. bends in the profile); we will discuss this further in Section 7.2.

\begin{figure}
\plotone{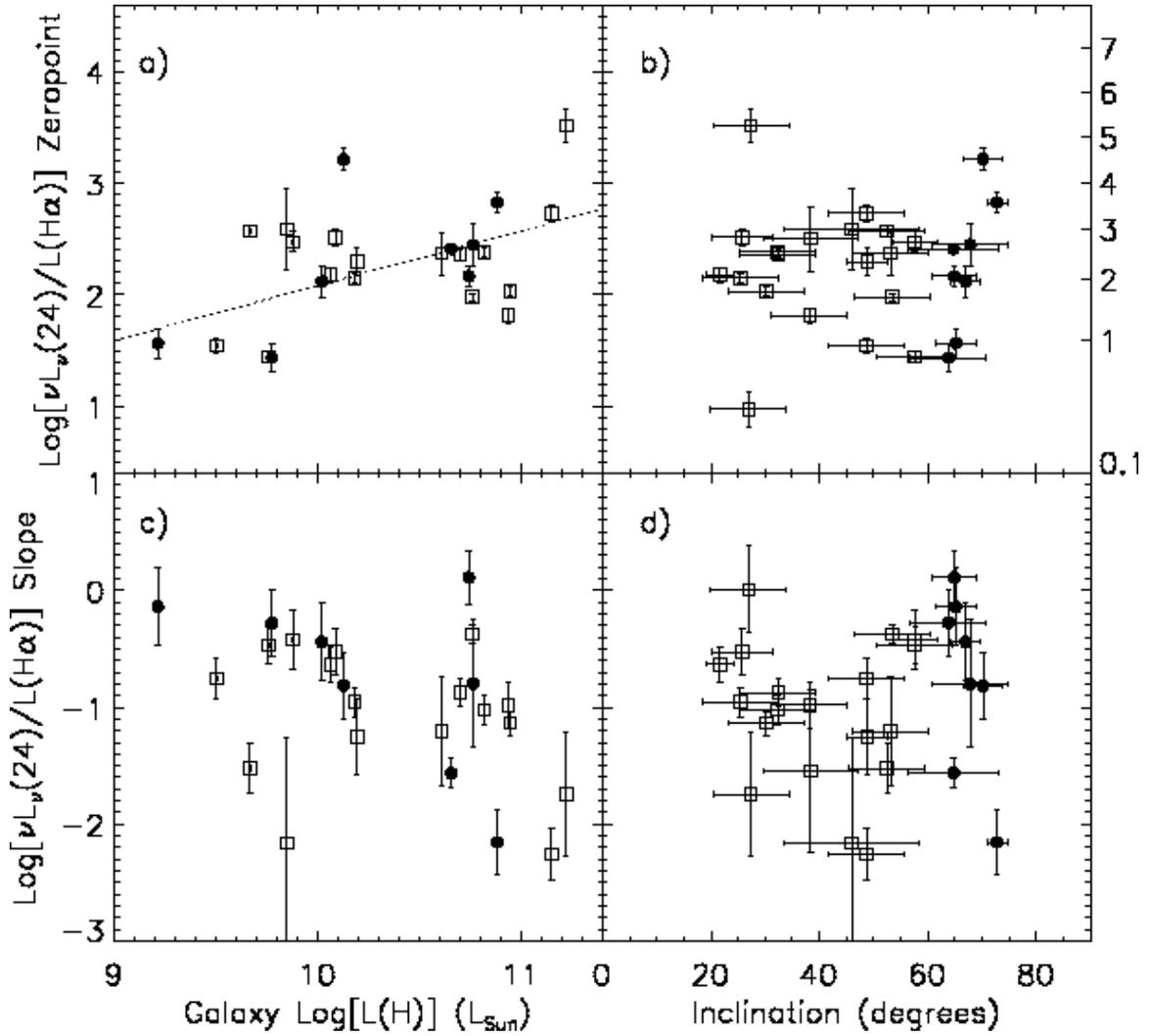}
\caption{Zeropoint and slope measurements derived from radial trends in star-forming region attenuation 
versus the total galaxy H-band luminosity and galaxy inclination.  The open squares correspond 
to low inclination ($i < 60^{\circ}$) galaxies, while the filled circles are more highly-inclined cases.  H-band 
luminosity errors are negligible; inclination errors are described in Section 2.  Slope 
and zeropoint error bars are formal fitting errors.  No strong correlations are seen; a tentative trend appears to exist between 
galaxy $H$-band luminosity and attenuation zeropoint and is corroborated by a similar trend between M$_{B}$ and zeropoint.  
Inclination does not play a dominant role in governing the observed slope and zeropoint of the attenuation profile.
}  
\label{fig:igallumfit}
\end{figure}

In using the empirical conversion outlined in Section 5, we are making the assumption 
that attenuation is the only mechanism responsible for varying the 24$\mu$m/H$\alpha$ ratio.  
We will proceed with this assumption here but will return to the question of what else could be 
causing the observed radial gradients in Section 7.2.
Using an unweighted, linear least-squares fit to the radial profiles we have investigated the relationship between 
these gradients and global galaxy properties of luminosity and inclination.  
We use the zeropoint, the linear extrapolation of the attenuation at the galaxy center, as a typical 
attenuation measure for each galaxy, and in Figure~\ref{fig:igallumfit} we plot the best fit attenuation slope and zeropoint against 
galaxy $H$-band luminosity \citep{jar03} and inclination.  
We find no strong trends; a tentative correlation appears to exist between the attenuation zeropoint and the galaxy $H$-band 
luminosity and is corroborated by a similar trend between M$_{B}$ and zeropoint.  
\citet{dal07} studied the global total infrared-to-ultraviolet (TIR/UV) ratios, often used as a proxy for attenuation, 
within the entire SINGS sample 
and found that TIR/UV increases with increasing $H$-band luminosity.  Our data are consistent with this result, 
though the relationship between attenuation zeropoint and galaxy $H$-band luminosity is not as strong.  
The galaxy inclination does not appear to regulate either the observed slope or zeropoint of the attenuation profile, 
consistent with findings by \citet{dal07} on global TIR/UV ratios. 
In terms of morphology, we see a weak tendency for early type galaxies with bars to show somewhat steeper slopes that 
early type galaxies without bars; no such relation is seen for late types.  We note that most of the 
highly-obscured regions in our sample are found in barred galaxies, consistent with other studies that 
have found an association between bars and circumnuclear star-forming regions \citep{reg06, ken98}.  

In the two overlapping cases between this work and \citet{boi04} (NGC~5194 and NGC~3031) we find similar results.  \citet{boi04} 
used FUV/IR ratios to derive UV attenuation estimates and compared these with A$_{H\alpha}$ measured from Balmer 
decrements of H~\textsc{ii} regions.  
In NGC~5194, they see a radial gradient of roughly 1 magnitude in H$\alpha$ attenuation over the optical radius of the galaxy, compared with 
$\sim$0.9 magnitudes in our study.  
In NGC~3031, they see a much shallower attenuation profile with a radial drop of $\sim$0.1 magnitudes, comparable to our result 
(A$_{H\alpha}\sim$0.2 magnitudes).

\begin{figure}
\plotone{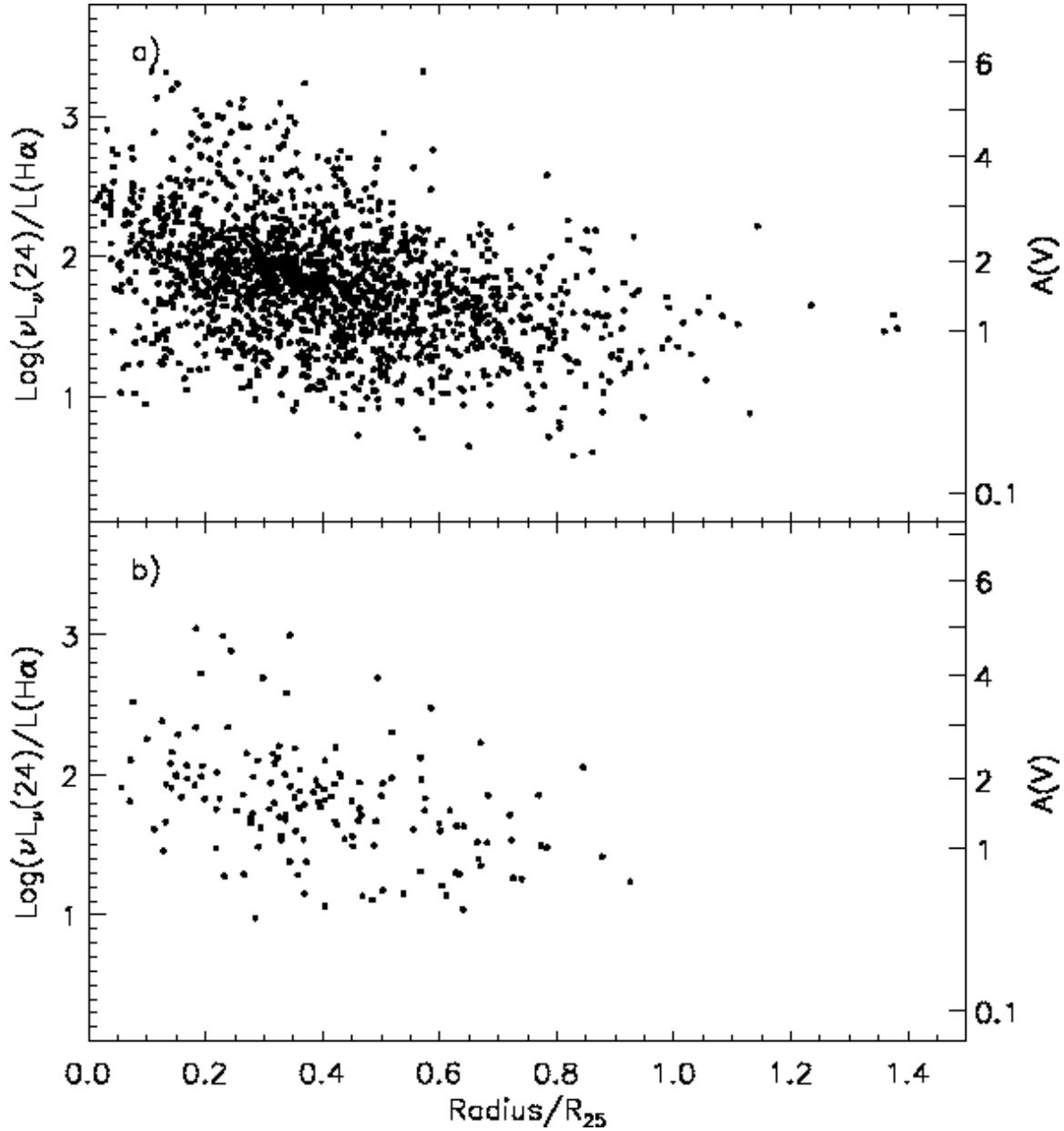}
\caption{(a) Radial profile of Log($\nu L_{\nu}(24)/L($H$\alpha)$) for the entire region sample.  While \citet{zar94} Figure 7c showed little indication of 
a radial trend in A$_{V}$, with our larger sample a radial trend is evident though much less tight than in any 
individual galaxy.  (b) Radial profile of Log($\nu L_{\nu}(24)/L($H$\alpha)$) for subset of our sample chosen to roughly replicate 
that of \citet{zar94} with only a tentative indication of a radial trend.  
The apparent lack of a radial attenuation gradient in \citet{zar94} was simply a consequence of plotting a region sample 
drawn from many galaxies, each with its own attenuation zeropoint.}  
\label{fig:zaritcompare}
\end{figure}

\citet{zar94} looked at the subject of attenuation gradients using optical spectra of a sample of 159 H~\textsc{ii} regions from 14 galaxies.  
They found only weak evidence for a radial trend in attenuation (see their Figure 7c).  However, we can interpret this result within the context 
of our study by noting that due to their more limited sample, they plotted the A$_{V}$ of the entire region sample versus fractional 
isophotal radius.  Since we find that the attenuation zeropoint can vary from galaxy to galaxy, viewing the region sample as a whole 
tends to wash out a radial signature.  Figure~\ref{fig:zaritcompare}a shows our version of their Figure 7c; a radial gradient is evident, although it is much 
less clear than what is seen in any individual galaxy.  When we randomly select 159 regions from our sample to match the sample size of \citet{zar94}, 
we obtain a result that is qualitatively indistinguishable from \citet{zar94} Figure 7c (Figure~\ref{fig:zaritcompare}b).

In view of the range in attenuation zeropoints found within the sample and the fact that 
our targets display decreasing radial attenuation gradients, we can return to the question of the nature of the tail of the attenuation 
distribution.  In defining a global attenuation cut, we selected the central regions of galaxies with the highest attenuation 
zeropoints.  To visualize that this is indeed the case, we have included in Figures~\ref{fig:iradialfitlow}-\ref{fig:iradialfithigh2} a dotted line corresponding to the 
attenuation cut applied in Section 6.1.  Regions that fall above the dotted line are preferentially the central regions of galaxies 
with overall high attenuation.  
This discussion underscores the fact that the highly-obscured fraction we quoted initially is quite arbitrary and dependent on the 
particular galaxy sample.  Nonetheless, while the precise value of the highly-obscured fraction is dependent on the sample, the statement that only 
$\sim$4\% of the regions in this particular set of galaxies have 24$\mu$m/H$\alpha$ attenuations higher than 2.7 (A$_{H\alpha} \sim$ 3.3) is 
representative of the robust result that these galaxies are not hosting a large number of 500~pc regions in which the vast 
majority of luminous stars are highly obscured.    

In a few galaxies there are a handful of regions that stand out relative to the rest of the regions in the host galaxy.  These regions are 
not necessarily highly-obscured in a global sense -- that is, they are not part of the highly-obscured tail of the global attenuation distribution -- but 
they do have higher attenuations than the rest of the region population of the host galaxy near that radius.  NGC~3938 and NGC~5474 are examples, 
with a few of these local outlier regions at 0.6-0.8 R$_{25}$.  Future work will be needed to look into the question of whether these regions 
are in some way exceptional or whether they simply reflect the stochasticity of the disk at large radii.

\section{DISCUSSION}

\subsection{Obscured Star Forming Regions}
The main purpose of this study was to characterize the attenuation distribution of infrared-selected star-forming regions as measured on large scales.  
In other words, the goal was to determine what fraction of bright infrared sources have H$\alpha$ counterparts and what fraction could be 
defined as highly obscured.  The median attenuation we found (A$_{H\alpha}\approx$1.4 magnitudes) indicates that roughly 72\% of 
the H$\alpha$ emission is typically ``missing'' 
from a 500~pc region, but cases where the H$\alpha$ is attenuated by more than 3 magnitudes are quite rare ($\sim$4\% of our sample).  
A complementary picture has been seen in direct measurements of mid-infrared silicate absorption within 
kiloparsec-sized regions in normal galaxies where most do not show foreground dust screen attenuation 
greater than $A_{V} \sim 3$ \citep{smi06}.  Similarly, optical nuclear spectra of SINGS galaxies show an average 
$A_{V} \sim 1$ \citep{dal06}. 
However, the fact that we see such a small fraction of uniformly highly-obscured regions could be due to a number of possible scenarios.  

\subsubsection{Timescales}
While dramatic examples of highly-obscured sources do exist in the local Universe on a variety of size scales, the main result 
of this study is that the fraction of such regions in normal, star-forming disk galaxies is small on 500~pc scales.  
Equivalently, we can say that the vast majority of current star formation in these galaxies is not 
hidden from view at optical wavelengths.  If we were to make the very simple assumptions that all large star-forming regions go through a highly-embedded 
phase and that a single star cluster dominates on 500~pc scales, we can estimate the lifetime of such an obscured period.  
Taking a typical H~\textsc{ii} region lifetime to be 
5-10 Myr, our observed obscured fraction of $\sim$4\% implies that a region would spend only $\sim10^{5}$ years 
being obscured by more than 3.3 magnitudes in H$\alpha$ ($A_{V} \gtrsim 4$ magnitudes).  
One could also account for these statistics if only a small fraction of the star-forming regions
undergo a more lengthy obscured phase.
These short time scales are in rough agreement with
those inferred for the lifetime and dynamical timescales of Young Stellar Object (YSO) outflows 
in nearby Galactic molecular clouds, \citep[e.g.,][]{qui05, zha05, bon96}.  
Furthermore, our sample consists of large 500~pc regions 
perhaps containing hundreds of stars which should shorten the time 
for sufficient cloud disruption even further.  

That said, in the above estimate we have envisioned simplified regions in which 
all stars are uniformly obscured.  In reality, each 500~pc region potentially encompasses  
many highly-obscured, compact infrared knots as well as emerged clusters which, 
when blended together in the large beam, would yield a lower average attenuation.  
We will return to this issue in the following section, but we note here that the 
observed highly-obscured fraction could also be explained by envisioning a toy model in which 
each 500~pc region half of the massive stars are embedded while half 
have broken out of one side of the giant molecular cloud in which they formed.  
Roughly half of the emerged cases have broken out on the opposite side 
of the cloud, and thus remain obscured from the point of view of an observer 
over the short lifetime of massive stars.  
The net H$\alpha$ absorption in this case would be of order 75\% (A$_{H\alpha}=1.5$), similar 
to the measured median attenuation in our sample.  This suggests breakout times 
of order one half of the massive star lifetime, or roughly 1 Myr.  
In either scenario, the small fraction of regions with nearly 
all of their stars highly-obscured indicates that breakout times must be less 
than the typical lifetime of a star-forming region and also less than or of 
order of the main sequence lifetime of the dominant sources of luminosity.

Although we have shown that infrared-luminous star-forming regions
with relatively faint optical counterparts are very rare in the 
extended disks of the SINGS spirals, we have also seen that they
are considerably more common in the circumnuclear regions of the
galaxies.  This difference may reflect the higher interstellar
pressures in the nuclear regions, which will act to slow the 
expansion of H~\textsc{ii} regions and prolong the lifetime of any dust-embedded
phase, as modeled, for example, by \citet{dop05}.  These 
regions are also characterized by much larger mean gas column
densities, however, so we would expect higher dust opacities even
without the dynamical effects of the higher pressure environment.
More detailed studies of individual regions will be needed to
separate these effects.

The strong association that we find between 24~$\mu$m sources and optical
H~\textsc{ii} regions also provides interesting contraints on the maximum lifetimes
of the star forming clouds.  If the molecular/atomic clouds and their
associated dust remained intact for lifetimes much longer than that
of the H~\textsc{ii} regions, we would expect many of the 24$\mu$m-bright H~\textsc{ii} regions
to be associated with the older clusters and OB associations, coincident
with ultraviolet-bright regions.  Since the dust can be heated by
non-ionizing stars such regions would stand out as having anomalously
high 24$\mu$m/H$\alpha$ ratios, effectively masquerading as highly
obscured regions.  While the obscured fraction that we find ($\sim$4\%) 
is therefore an upper limit, the scarcity of regions with high 
24$\mu$m/H$\alpha$ ratios means that 24$\mu$m-bright 
clusters older than 5--10 Myr are even more rare.  
This suggests that the dust and gas clouds surrounding
the star clusters/associations must largely dissipate on this time scale.
Independent evidence for such time scales comes from observations of
expansion lifetimes of HI giant shells in the LMC and SMC, which
also show an absence of objects with expansion ages of more than 10 Myr \citep[e.g.,][]{kim99, hat05}.

\subsubsection{Size scales}
Our results do not preclude the presence of embedded protostars or ultra-compact H~\textsc{ii} regions within the larger complexes that we have studied.
In this work we have presented attenuation measurements for large, 500~pc apertures.  A common aperture size allowed us to compare the same 
spatial scale between galaxies at distances ranging from 3 to 25 Mpc; however, a 500~pc aperture corresponds to a larger scale than that 
of a typical H~\textsc{ii} region, let alone a compact H~\textsc{ii} region ($\sim$1~pc).  In every aperture applied to the 24~$\mu$m data, we are 
potentially summing together a number of smaller infrared star-forming knots with varying attenuations; the net result of this blending is that the 
attenuation measured is an average for the region.  This situation is of course only exacerbated in global star formation 
studies at high redshift in which measurements are necessarily averaged over an entire galaxy.  
In Section 6.1, we addressed the question of whether probing to smaller scales 
would have drastically changed the fraction of highly obscured sources.  Based on these tests we do not expect our results on the highly-obscured 
fraction to change dramatically down to 100~pc scales.  When we repeat our analysis on similar datasets for the Antennae galaxies, for example, 
we are able recover the highly-obscured case despite the blending that occurs due to our large aperture.  The region itself is $\sim$50~pc with 
an attenuation of $\sim$9-10 magnitudes in V derived from NIR spectra \citep{gil00}; our empirically-calibrated photometric attenuation measurement 
is $\sim$4.8 in V ($A_{H\alpha}\approx$ 3.9) using a 500~pc aperture, with the region being a clear (8$\sigma$) outlier from the rest of the regions in the system.  
Thus, if a large sub-population of similarly luminous and embedded regions existed in our sample of SINGS galaxies, 
we should have found them in this study.  

This is not to suggest the absence of much smaller or less luminous embedded infrared sources.  We know from studies of the Milky Way 
that ultracompact and compact H~\textsc{ii} regions exist on scales of $\sim$0.1 - 1~pc with TIR luminosities of $\sim10^{37} - 10^{39}$ 
erg s$^{-1}$ \citep{cro05,aka05}.  Star-forming regions in our sample are typically at least $\sim$100 times more luminous than a compact 
H~\textsc{ii} region just in the 24$\mu$m-band ($\sim10^{39}-10^{42}$ erg s$^{-1}$).  
No doubt a multitude of such regions exist within each galaxy in our sample, but they are not large enough or luminous enough to be studied with the 
current data.

\subsection{Radial Gradients}

The fact that we see few highly-obscured cases and such smooth radial gradients suggests that the measured attenuations on these scales 
are dictated to a greater extent by the global properties of the galaxy such as metallicity, gas surface density, inclination, geometry, 
or gradual changes in the region SED than by region-to-region fluctuations.  
Since metallicity is related to dust content and since decreasing gradients in metallicity are observed in many galaxies, it is reasonable 
to expect it to contribute to the attenuation properties of star-forming regions in a manner that varies with radius \citep{zar94}.  
In contrast to previous work \citep{boi04, boi05}, a preliminary comparison of the metallicity and attenuation gradients for the low 
inclination galaxies in our sample did not suggest any correlation between the two.  However, this is perhaps not surprising as our sample 
does not cover a wide baseline in metallicity, and   
gradients in gas column density in galaxies may play a larger role in the global attenuation profiles.  
\citet{rou05} found a similar radial gradient in FUV/TIR in NGC~300 and noted that while the dispersion in FUV/TIR 
ratios of individual regions was at least a factor of $\sim$2 ($\gtrsim$0.30 dex), the standard deviation of abundances was only 0.11 dex.  
From this they suggested that metallicity was likely not the only factor responsible for the gradient.  
The inclination of the galaxy is a consideration since distributed dust seen along a line-of-sight through the disk may cause additional 
attenuation of the region's H$\alpha$ flux.  In this sample, however, galaxy inclination does not appear to play a dominant role in 
regulating either the slope or the zeropoint of the global attenuation profile.  
Instead, the situation appears to be more complicated and may depend strongly on properties of an individual galaxy such as the thickness and 
morphology of dust lanes and the dust and gas scale heights.  

It is interesting to note that while the majority of the sample show linear radial trends in 24$\mu$m/H$\alpha$, there are two cases (NGC~1097 and NGC~4254) that 
show a bend in the profile at roughly 0.6 R$_{25}$.  Since metallicity presumably plays a role in the dust-to-gas ratio, 
it is reasonable to ask whether a bend in the attenuation profile could be related to a bend in the abundance profile of a galaxy.  
Previous work by numerous authors presented examples of bends in abundance gradients and discussed the question of whether they were real 
or simply an artifact of the method used to determine (O/H) \citep[ and references therein]{zar94}.  
\citet{pily03} argued strongly that the bends were in fact artificial, that they were due to either an inappropriate application to low metallicity regions 
of the high-metallicity relationship between oxygen abundace and strong lines or due to a gradient in the excitation parameter within the disk.  
While beyond the scope of this paper, it is interesting to note that we find examples of bends in attenuation gradients using a completely 
different technique.  This leads to the question of whether these bends could be related to gradients in the [N~\textsc{ii}]/H$\alpha$ ratio or 
systematics inherent to the empirical method of estimating attenuation, or whether they represent a real change in the attenuation properties of the disk.

The results summarized above assume that all of the radial trends observed
in the 24$\mu$m/H$\alpha$ ratios of the H~\textsc{ii} regions arise from changes
in dust attenuation.  However it
is possible that some of the radial trends that we observe are produced
by systematic changes in the infrared SED of the emitting regions with galactocentric radius, 
independent of extinction.  For example, if
the diameter of H~\textsc{ii} regions becomes systematically larger with galactocentric
radius, this would lower the characteristic intensity of the heating
radiation field at larger radii and tend to lower the mean dust temperatures
and the fraction of total dust luminosity emitted in the 24-$\mu$m band.
Spatially-resolved measurements of the infrared spectral energy distributions
of SINGS galaxies show little evidence for such radial gradients in infrared SED
shape \citep[][]{ben06, dal05, per06}, 
but we cannot exclude the possibility that there is at least some small contribution
from this effect to the gradients that we observe in Figures~\ref{fig:iradialfitlow}-\ref{fig:iradialfithigh2}.

\section{CONCLUSIONS}
In this work, we have studied the incidence of obscured star formation and the distribution of attenuation properties for 
infrared-selected star-forming regions measured on large scales ($\sim$500~pc) in a sample of 38 SINGS galaxies.  
We have made use of an empirical method for deriving H$\alpha$ 
attenuations which combines the 24~$\mu$m and H$\alpha$ datasets \citep{ken07a}.  Under the assumption that the 24$\mu$m/H$\alpha$ ratio 
can be used as a proxy for attenuation, our main results are as follows:

\begin{enumerate} 
\item  We do {\it not} find a distinct sub-population of uniformly highly-obscured regions on large ($\sim$500~pc) scales.  Instead our data suggest a 
continuous distribution of attenuations (power law index of $\sim$1.4) within the entire sample of 1772 regions.  

\item  With a few exceptions, those regions that do have higher obscuration (A$_{H\alpha} > 3.3$) are the bright, central regions of luminous galaxies with high 
global attenuation.  The regions in the tail of the attenuation distribution represent a small fraction ($\sim$4\%) of the region population. 

\item  Most galaxies show a decreasing radial trend in 24$\mu$m/H$\alpha$; the current sample shows a range of slopes and zeropoints.  
A weak correlation may exist between the attenuation zeropoint and the galaxy luminosity, consistent with previous studies, but 
the galaxy inclination does not appear to play a dominant role in either the observed slope or zeropoint of the profile.

\item This study reports on the attenuation (A$_{H\alpha}$) properties of large ($\sim$500~pc) star-forming regions, but does 
not comment on the prevalence of embedded infrared sources that no doubt exist on smaller scales and at lower 
luminosities similar to compact and ultracompact H{\sc ii} regions in the Milky Way.  

\item A more general conclusion of our work is the high degree of spatial correlation between 24~$\mu$m and H$\alpha$ emission in 
SINGS galaxies.  In agreement with other studies \citep[e.g.,][]{per06}, this work suggests that for every bright 24~$\mu$m source 
in a normal galaxy there is a visible H$\alpha$ counterpart, a fact which opens the door not only to attenuation studies but also to the 
possibility of using the empirical [24$\mu$m+H$\alpha$] combination as a new measure of star-formation rates in galaxies.  

\end{enumerate}

\acknowledgments

We are grateful to Kristian Finlator and Christy Tremonti 
for many thoughtful discussions.  M. P. was supported by a NSF Graduate Research Fellowship.  

This work is based in part on observations made with the Spitzer Space Telescope, which is operated by the Jet Propulsion Laboratory, 
California Institute of Technology under a contract with NASA.  Support for this work was provided by NASA through an award issued by JPL/Caltech.

This research has made use of the NASA/IPAC Extragalactic Database (NED) which is operated by the Jet Propulsion Laboratory, 
California Institute of Technology, under contract with the National Aeronautics and Space Administration.

This publication makes use of data products from the Two Micron All Sky Survey, which is a joint project of the University of Massachusetts and 
the Infrared Processing and Analysis Center/California Institute of Technology, funded by the National Aeronautics and Space Administration 
and the National Science Foundation.

\clearpage
\input{tab1.tex}

\input{tab2.tex}

\input{tab3.tex}

\clearpage



\clearpage


\end{document}

%% file: tab1.tex
\begin{deluxetable}{cccccccccccc}
\tabletypesize{\scriptsize}
\tablecaption{Galaxy Sample}
\tablewidth{0pt}
\tablehead{
\colhead{Galaxy} & \colhead{Distance} & \colhead{M$_{B}$} & \colhead{D$_{25}$} & \colhead{Type} & \colhead{{\it i}} & \colhead{N$_{Reg}$} & \colhead{Log[$\nu L_{\nu}$(24)]} & \colhead{N$_{HO}$} & Slope & Zeropoint \\
 & & & & & & & Limit & & & \\
(1) & (2) & (3) & (4) & (5) & (6) & (7) & (8) & (9) & (10) & (11) }
\startdata
                NGC 0024 &      7.3$\pm$2.3 (i) &   -17.14$\pm$0.13 &      5.8$\times$ 1.3 &                   SA(s)c &       84$\pm$ 7 &        7 &    39.17$\pm$38.52 &        0 &    - &    - \\ 
                NGC 0337 &     22.4$\pm$2.3 (i) &   -19.69$\pm$0.13 &      2.9$\times$ 1.8 &                   SB(s)d &       53$\pm$ 7 &       11 &    40.48$\pm$39.94 &        0 &    -1.21$\pm$0.47 &     2.37$\pm$0.19 \\ 
                NGC 0628 &      7.3$\pm$1.5 (k) &   -19.37$\pm$0.10 &     10.5$\times$ 9.5 &                   SA(s)c &       26$\pm$ 6 &       70 &    39.46$\pm$38.40 &        2 &    -0.53$\pm$0.20 &     2.51$\pm$0.07 \\ 
                NGC 0925 &      9.1$\pm$0.2 (b) &   -19.11$\pm$0.11 &     10.5$\times$ 5.9 &                  SAB(s)d &       58$\pm$ 4 &       35 &    39.24$\pm$37.95 &        1 &    -0.42$\pm$0.25 &     2.47$\pm$0.09 \\ 
                NGC 1097 &     17.1$\pm$2.3 (i) &   -20.93$\pm$0.07 &      9.3$\times$ 6.3 &                (R)SB(r)b &       49$\pm$ 7 &       66 &    39.50$\pm$38.29 &        3 &    -2.25$\pm$0.22 &     2.73$\pm$0.08 \\ 
                NGC 1291 &     10.8$\pm$2.3 (i) &   -20.78$\pm$0.04 &      9.8$\times$ 8.1 &              (R)SB(l)0/a &       35$\pm$ 7 &        1 &    39.30$\pm$37.52 &        0 &    - &    - \\ 
                NGC 1316 &     24.3$\pm$2.3 (i) &   -22.51$\pm$0.08 &     12.0$\times$ 8.5 &               (R)SAB(s)0 &       46$\pm$ 7 &        1 &    40.94$\pm$40.10 &        0 &    - &    - \\ 
                NGC 1512 &     11.8$\pm$2.3 (i) &   -19.23$\pm$0.10 &      8.9$\times$ 5.6 &                  SB(r)ab &       52$\pm$ 7 &        9 &    39.77$\pm$38.87 &        0 &    -1.52$\pm$0.21 &     2.57$\pm$0.05 \\ 
                NGC 1566 &     20.3$\pm$2.3 (i) &   -21.21$\pm$0.03 &      8.3$\times$ 6.6 &             (R)SAB(rs)bc &       38$\pm$ 7 &       62 &    39.61$\pm$39.17 &        0 &    -0.98$\pm$0.20 &     1.81$\pm$0.07 \\ 
                NGC 1705 &      5.1$\pm$0.6 (m) &   -15.77$\pm$0.13 &      1.9$\times$ 1.4 &                  SA0 pec &       44$\pm$ 7 &        2 &    39.77$\pm$38.91 &        0 &    - &    - \\ 
                NGC 2403 &      3.2$\pm$0.2 (b) &   -18.61$\pm$0.07 &     21.9$\times$12.3 &                 SAB(s)cd &       58$\pm$ 7 &       88 &    38.36$\pm$37.18 &        0 &    -0.47$\pm$0.16 &     1.45$\pm$0.05 \\ 
                NGC 2841 &     14.1$\pm$1.5 (h) &   -20.66$\pm$0.10 &      8.1$\times$ 3.5 &                   SA(r)b &       67$\pm$ 1 &       28 &    39.53$\pm$38.87 &        0 &    -0.44$\pm$0.33 &     2.12$\pm$0.14 \\ 
                NGC 2976 &      3.6$\pm$0.4 (d) &   -16.94$\pm$0.13 &      5.9$\times$ 2.7 &                  SAc pec &       65$\pm$ 3 &       14 &    39.51$\pm$39.18 &        0 &    -0.14$\pm$0.33 &     1.56$\pm$0.13 \\ 
                NGC 3031 &      3.6$\pm$0.1 (b) &   -19.91$\pm$0.03 &     29.6$\times$14.1 &                  SA(s)ab &       64$\pm$ 4 &       81 &    38.64$\pm$37.32 &        0 &    -0.28$\pm$0.29 &     1.43$\pm$0.13 \\ 
                NGC 3184 &     11.1$\pm$2.0 (g) &   -19.87$\pm$0.10 &      7.4$\times$ 6.9 &                SAB(rs)cd &       22$\pm$ 3 &       80 &    39.40$\pm$38.77 &        0 &    -0.63$\pm$0.15 &     2.18$\pm$0.06 \\ 
                NGC 3190 &     20.9$\pm$2.3 (i) &   -19.48$\pm$0.10 &      4.4$\times$ 1.5 &               SA(s)a pec &       90$\pm$ 7 &        5 &    40.57$\pm$39.71 &        0 &    - &    - \\ 
                NGC 3198 &     13.7$\pm$0.5 (b) &   -19.81$\pm$0.10 &      8.5$\times$ 3.3 &                  SB(rs)c &       70$\pm$ 3 &       43 &    39.50$\pm$38.28 &       35 &    -0.82$\pm$0.28 &     3.22$\pm$0.10 \\ 
                 IC 2574 &      4.0$\pm$0.4 (e) &   -17.22$\pm$0.19 &     13.2$\times$ 5.4 &                  SAB(s)m &       69$\pm$ 3 &       13 &    38.59$\pm$37.43 &        0 &    - &    - \\ 
                NGC 3351 &      9.3$\pm$0.4 (b) &   -19.32$\pm$0.10 &      7.4$\times$ 5.0 &                   SB(r)b &       49$\pm$ 4 &       13 &    39.46$\pm$38.79 &        0 &    -1.25$\pm$0.32 &     2.30$\pm$0.12 \\ 
                NGC 3521 &     10.1$\pm$2.3 (i) &   -20.18$\pm$0.10 &     11.0$\times$ 5.1 &                SAB(rs)bc &       65$\pm$ 8 &      100 &    39.61$\pm$39.08 &        0 &    -1.56$\pm$0.13 &     2.41$\pm$0.04 \\ 
                NGC 3627 &      9.4$\pm$0.1 (b) &   -20.21$\pm$0.13 &      9.1$\times$ 4.2 &                  SAB(s)b &       65$\pm$ 4 &       71 &    39.57$\pm$39.26 &        2 &     0.11$\pm$0.23 &     2.17$\pm$0.09 \\ 
                NGC 3938 &     13.3$\pm$2.3 (i) &   -19.73$\pm$0.10 &      5.4$\times$ 4.9 &                   SA(s)c &       25$\pm$ 7 &       77 &    39.46$\pm$38.66 &        0 &    -0.96$\pm$0.12 &     2.15$\pm$0.04 \\ 
                NGC 4254 &     16.6$\pm$2.0 (i*) &   -20.66$\pm$0.08 &      5.4$\times$ 4.7 &                   SA(s)c &       30$\pm$ 7 &      119 &    39.69$\pm$39.19 &        0 &    -1.13$\pm$0.10 &     2.03$\pm$0.04 \\ 
                NGC 4321 &     14.3$\pm$0.5 (l) &   -20.73$\pm$0.08 &      7.4$\times$ 6.3 &                 SAB(s)bc &       32$\pm$ 7 &       87 &    39.64$\pm$39.40 &        0 &    -0.87$\pm$0.13 &     2.36$\pm$0.05 \\ 
                NGC 4450 &     16.6$\pm$2.0 (i*) &   -20.20$\pm$0.08 &      5.2$\times$ 3.9 &                  SA(s)ab &       42$\pm$ 7 &        3 &    39.89$\pm$39.03 &        0 &    - &    - \\ 
                NGC 4536 &     14.4$\pm$0.3 (b) &   -19.64$\pm$0.08 &      7.6$\times$ 3.2 &                SAB(rs)bc &       68$\pm$ 7 &       25 &    39.70$\pm$38.84 &        1 &    -0.80$\pm$0.54 &     2.45$\pm$0.19 \\ 
                NGC 4631 &      8.1$\pm$2.3 (i) &   -19.78$\pm$0.16 &     15.5$\times$ 2.7 &                   SB(s)d &       90$\pm$ 7 &       68 &    39.59$\pm$39.00 &        1 &    - &    - \\ 
                NGC 4725 &     11.9$\pm$0.3 (b) &   -20.27$\pm$0.13 &     10.7$\times$ 7.6 &             SAB(r)ab pec &       46$\pm$ 7 &       32 &    39.44$\pm$38.96 &        0 &    -2.16$\pm$0.89 &     2.58$\pm$0.36 \\ 
                NGC 5194 &      7.8$\pm$2.3 (i) &   -20.49$\pm$0.06 &     11.2$\times$ 6.9 &              SA(s)bc pec &       54$\pm$12 &      217 &    39.33$\pm$39.08 &        0 &    -0.38$\pm$0.08 &     1.98$\pm$0.03 \\ 
               Tololo 89 &     16.7$\pm$2.3 (i) &   -18.33$\pm$0.17 &      2.8$\times$ 1.7 &                 SB(rs)dm &       54$\pm$ 7 &        3 &    39.96$\pm$38.83 &        0 &    - &    - \\ 
                NGC 5474 &      6.8$\pm$1.4 (a) &   -17.88$\pm$0.15 &      4.8$\times$ 4.3 &              SA(s)cd pec &       27$\pm$ 7 &       15 &    38.90$\pm$37.78 &        0 &     0.01$\pm$0.37 &     0.97$\pm$0.16 \\ 
                NGC 5713 &     29.4$\pm$2.3 (i) &   -20.50$\pm$0.15 &      2.8$\times$ 2.5 &            SAB(rs)bc pec &       27$\pm$ 7 &       14 &    40.68$\pm$40.41 &       13 &    -1.74$\pm$0.53 &     3.52$\pm$0.16 \\ 
                 IC 4710 &      9.0$\pm$2.3 (i) &   -17.26$\pm$0.20 &      3.6$\times$ 2.8 &                   SB(s)m &       40$\pm$ 7 &        5 &    39.44$\pm$38.37 &        0 &    - &    - \\ 
                NGC 6822 &      0.5$\pm$0.0 (j) &   -14.03$\pm$0.06 &     15.5$\times$13.5 &                   IB(s)m &        0$\pm$ 7 &       13 &    37.70$\pm$37.45 &        0 &    - &    - \\ 
                NGC 6946 &      6.8$\pm$1.4 (c) &   -19.55$\pm$0.10 &     11.5$\times$ 9.8 &                SAB(rs)cd &       32$\pm$ 9 &      158 &    39.51$\pm$38.94 &        4 &    -1.02$\pm$0.13 &     2.38$\pm$0.05 \\ 
                NGC 7331 &     14.5$\pm$0.6 (b) &   -20.46$\pm$0.10 &     10.5$\times$ 3.7 &                   SA(s)b &       73$\pm$ 2 &       77 &    39.86$\pm$38.78 &        7 &    -2.15$\pm$0.27 &     2.83$\pm$0.09 \\ 
                NGC 7552 &     21.0$\pm$2.3 (i) &   -20.36$\pm$0.13 &      3.4$\times$ 2.7 &               (R)SB(s)ab &       38$\pm$ 7 &       11 &    40.47$\pm$40.05 &        0 &    -1.55$\pm$0.70 &     2.50$\pm$0.29 \\ 
                NGC 7793 &      3.8$\pm$0.4 (f) &   -18.28$\pm$0.05 &      9.3$\times$ 6.3 &                   SA(s)d &       49$\pm$ 3 &       48 &    39.14$\pm$37.80 &        0 &    -0.76$\pm$0.17 &     1.55$\pm$0.07 \\ 
\enddata
\tablecomments{Col.~(1) Galaxy name; Col.~(2) Distance~(Mpc) with associated reference; Col.~(3) Absolute B magnitude from \citet{devauc91} (RC3); Col.~(4) Major and minor axis diameters at the 25 magnitude isophote from RC3 (arcmin); Col.~(5) Morphological type from RC3; Col.~(6) Inclination derived using eq.~\ref{eqn:incl} (degrees); Col.~(7) Number of infrared-selected regions included in this study; Col.~(8) Log of the 24$\mu$m luminosity of the faintest region (erg s$^{-1}$); Col.~(9) Number of highly obscured regions, as defined in Section 6.1; Col~(10) Slope of radial attenuation gradient as defined in Section 6.4 (dex/kpc); Col~(11) Zeropoint of radial attenuation gradient as defined in Section 6.4 (dex).}
\tablenotetext{*}{Distance set to Virgo Cluster center.}
\tablerefs{(a) \citet{dk00}; (b) \citet{free01}; (c) \citet{kar00}; (d) \citet{kar02}; (e) \citet{kar03a}; (f) \citet{kar03b}; (g) \citet{leo02}; (h) \citet{mac01}; (i) \citet{mast07}; (j) \citet{piet04}; (k) \citet{shap01}; (l) \citet{ton01}; (m) \citet{tosi01}}
\end{deluxetable}

%% file: tab2.tex
\begin{deluxetable}{ccc}
\tabletypesize{\scriptsize}
\tablecaption{H$\alpha$ Correction Factors}
\tablewidth{0pt}
\tablehead{
\colhead{} & \colhead{[N~\textsc{ii}]~$\lambda$6584/H$\alpha$} & \colhead{Final Correction} \\
Galaxy & Ratio\tablenotemark{a} & Factor}
\startdata
                NGC 0024 &    0.268$\pm$0.008 &   0.8947$\pm$0.0028 \\ 
                NGC 0337 &    0.197$\pm$0.002 &   0.9819$\pm$0.0002 \\ 
                NGC 0628 &    0.497$\pm$0.014 &   0.9645$\pm$0.0009 \\ 
                NGC 0925 &    0.233$\pm$0.006 &   0.7814$\pm$0.0035 \\ 
                NGC 1097 &    0.485$\pm$0.004 &   0.7323$\pm$0.0017 \\ 
                NGC 1291 &    1.060$\pm$0.157 &   0.9487$\pm$0.0051 \\ 
                NGC 1316 &    2.489$\pm$0.126 &   0.6646$\pm$0.0103 \\ 
                NGC 1512 &    0.481$\pm$0.012 &   0.9713$\pm$0.0005 \\ 
                NGC 1566 &    0.507$\pm$0.004 &   0.9737$\pm$0.0002 \\ 
                NGC 1705 &    0.065$\pm$0.002 &   0.9560$\pm$0.0012 \\ 
                NGC 2403 &    0.231$\pm$0.002 &   0.7679$\pm$0.0012 \\ 
                NGC 2841 &    0.866$\pm$0.027 &   0.4972$\pm$0.0059 \\ 
                NGC 2976 &    0.269$\pm$0.004 &   0.7424$\pm$0.0020 \\ 
                NGC 3031 &     0.36$\pm$0.30\tablenotemark{b} &     0.67$\pm$0.12 \\ 
                NGC 3184 &    0.439$\pm$0.009 &   0.6571$\pm$0.0036 \\ 
                NGC 3190 &    0.958$\pm$0.091 &   0.4594$\pm$0.0206 \\ 
                NGC 3198 &    0.348$\pm$0.008 &   0.7125$\pm$0.0036 \\ 
                 IC 2574 &     0.11$\pm$0.30\tablenotemark{b} &     0.87$\pm$0.18 \\ 
                NGC 3351 &    0.458$\pm$0.004 &   0.6292$\pm$0.0017 \\ 
                NGC 3521 &    0.426$\pm$0.004 &   0.6802$\pm$0.0016 \\ 
                NGC 3627 &    0.495$\pm$0.005 &   0.6381$\pm$0.0018 \\ 
                NGC 3938 &    0.310$\pm$0.009 &   0.7460$\pm$0.0040 \\ 
                NGC 4254 &    0.307$\pm$0.002 &   0.7096$\pm$0.0012 \\ 
                NGC 4321 &    0.396$\pm$0.005 &   0.8285$\pm$0.0014 \\ 
                NGC 4450 &    0.873$\pm$0.043 &   0.4777$\pm$0.0095 \\ 
                NGC 4536 &    0.432$\pm$0.005 &   0.8060$\pm$0.0015 \\ 
                NGC 4631 &     0.34$\pm$0.30\tablenotemark{b} &     0.71$\pm$0.12 \\ 
                NGC 4725 &    1.261$\pm$0.077 &   0.5627$\pm$0.0105 \\ 
                NGC 5194 &    0.524$\pm$0.004 &   0.6023$\pm$0.0013 \\ 
               Tololo 89 &     0.18$\pm$0.30\tablenotemark{b} &     0.90$\pm$0.09 \\ 
                NGC 5474 &     0.15$\pm$0.30\tablenotemark{b} &     0.83$\pm$0.17 \\ 
                NGC 5713 &    0.418$\pm$0.003 &   0.6565$\pm$0.0011 \\ 
                 IC 4710 &     0.12$\pm$0.30\tablenotemark{b} &     0.99$\pm$0.01 \\ 
                NGC 6822 &     0.03$\pm$0.30\tablenotemark{b} &     1.00$\pm$0.03 \\ 
                NGC 6946 &     0.31$\pm$0.30\tablenotemark{b} &     0.71$\pm$0.13 \\ 
                NGC 7331 &    0.507$\pm$0.007 &   0.6433$\pm$0.0025 \\ 
                NGC 7552 &     0.44$\pm$0.30\tablenotemark{b} &     0.97$\pm$0.01 \\ 
                NGC 7793 &    0.287$\pm$0.005 &   0.9822$\pm$0.0002 \\ 
\enddata
\tablecomments{The final correction factor multiplies the raw flux measurement and corrects for the [N~\textsc{ii}]~$\lambda\lambda$(6584,6548)/H$\alpha$ ratios as well as the H$\alpha$ filter transmission.  Quoted errors are formal measurement errors only and do not include physical uncertainties, e.g., due to variations in the [N~\textsc{ii}]~$\lambda\lambda$(6584,6548)/H$\alpha$ ratio within the galaxy.}
\tablenotetext{a}{[N~\textsc{ii}]~$\lambda\lambda$(6584,6548)/H$\alpha$ ratios were derived as discussed in Section 3.  The assumed [N~\textsc{ii}]~$\lambda$6548/[N~\textsc{ii}]~$\lambda$6584 ratio is 0.335, consistent with other studies \citep{cal05}.}
\tablenotetext{b}{Ratio derived using eq.~\ref{eqn:niihalpha}.  The standard deviation of this relation is estimated to be $\pm$0.30.}
\end{deluxetable}

%% file: tab3.tex
\begin{deluxetable}{ccccccc}
\tabletypesize{\scriptsize}
\tablecaption{Aperture Correction Factors}
\tablewidth{0pt}
\tablehead{
\colhead{Galaxy} & \colhead{H$\alpha$} & \colhead{3.6$\mu$m} & \colhead{4.5$\mu$m} & \colhead{5.8$\mu$m} & \colhead{8.0$\mu$m} & \colhead{24$\mu$m} }
\startdata
                NGC 0024 &     1.03 &     1.02 &     1.02 &     1.02 &     1.03 &     1.64 \\ 
                NGC 0337 &     4.46 &     4.94 &     4.94 &     5.18 &     5.69 &     5.85 \\ 
                NGC 0628 &     1.02 &     1.04 &     1.04 &     1.05 &     1.07 &     1.64 \\ 
                NGC 0925 &     1.09 &     1.07 &     1.07 &     1.07 &     1.08 &     1.72 \\ 
                NGC 1097 &     2.56 &     2.95 &     2.95 &     3.08 &     3.37 &     3.21 \\ 
                NGC 1291 &     1.16 &     1.23 &     1.23 &     1.25 &     1.31 &     1.88 \\ 
                NGC 1316 &     4.21 &     4.94 &     4.94 &     5.18 &     5.69 &     5.85 \\ 
                NGC 1512 &     1.47 &     1.61 &     1.61 &     1.67 &     1.78 &     2.21 \\ 
                NGC 1566 &     2.61 &     2.95 &     2.95 &     3.08 &     3.37 &     3.21 \\ 
                NGC 1705 &     1.00 &     1.00 &     1.00 &     1.00 &     1.00 &     1.34 \\ 
                NGC 2403 &     1.00 &     1.00 &     1.00 &     1.00 &     1.00 &     1.34 \\ 
                NGC 2841 &     1.55 &     1.61 &     1.61 &     1.67 &     1.78 &     2.21 \\ 
                NGC 2976 &     1.00 &     1.00 &     1.00 &     1.00 &     1.00 &     1.34 \\ 
                NGC 3031 &     1.00 &     1.00 &     1.00 &     1.00 &     1.00 &     1.34 \\ 
                NGC 3184 &     1.16 &     1.23 &     1.23 &     1.25 &     1.31 &     1.88 \\ 
                NGC 3190 &     4.41 &     4.94 &     4.94 &     5.18 &     5.69 &     5.85 \\ 
                NGC 3198 &     1.47 &     1.61 &     1.61 &     1.67 &     1.78 &     2.21 \\ 
                 IC 2574 &     1.00 &     1.00 &     1.00 &     1.00 &     1.00 &     1.34 \\ 
                NGC 3351 &     1.10 &     1.15 &     1.15 &     1.17 &     1.21 &     1.72 \\ 
                NGC 3521 &     1.16 &     1.23 &     1.23 &     1.25 &     1.31 &     1.88 \\ 
                NGC 3627 &     1.15 &     1.23 &     1.23 &     1.25 &     1.31 &     1.88 \\ 
                NGC 3938 &     1.49 &     1.61 &     1.61 &     1.67 &     1.78 &     2.21 \\ 
                NGC 4254 &     2.60 &     2.95 &     2.95 &     3.08 &     3.37 &     3.21 \\ 
                NGC 4321 &     1.47 &     1.61 &     1.61 &     1.67 &     1.78 &     2.21 \\ 
                NGC 4450 &     2.20 &     2.11 &     2.10 &     2.15 &     2.22 &     3.21 \\ 
                NGC 4536 &     1.48 &     1.61 &     1.61 &     1.67 &     1.78 &     2.21 \\ 
                NGC 4631 &     1.08 &     1.07 &     1.07 &     1.07 &     1.08 &     1.72 \\ 
                NGC 4725 &     1.49 &     1.61 &     1.61 &     1.67 &     1.78 &     2.21 \\ 
                NGC 5194 &     1.02 &     1.04 &     1.04 &     1.05 &     1.07 &     1.64 \\ 
               Tololo 89 &     2.63 &     2.95 &     2.95 &     3.08 &     3.37 &     3.21 \\ 
                NGC 5474 &     1.01 &     1.01 &     1.01 &     1.01 &     1.01 &     1.55 \\ 
                NGC 5713 &     4.26 &     4.94 &     4.94 &     5.18 &     5.69 &     5.85 \\ 
                 IC 4710 &     1.11 &     1.15 &     1.15 &     1.17 &     1.21 &     1.72 \\ 
                NGC 6822 &     1.00 &     1.00 &     1.00 &     1.00 &     1.00 &     1.34 \\ 
                NGC 6946 &     1.01 &     1.01 &     1.01 &     1.01 &     1.01 &     1.55 \\ 
                NGC 7331 &     1.49 &     1.61 &     1.61 &     1.67 &     1.78 &     2.21 \\ 
                NGC 7552 &     4.28 &     4.94 &     4.94 &     5.18 &     5.69 &     5.85 \\ 
                NGC 7793 &     1.00 &     1.00 &     1.00 &     1.00 &     1.00 &     1.34 \\ 
\enddata
\tablecomments{Raw flux measurements are multiplied by these aperture correction factors to produce the corrected values.}
\end{deluxetable}